\documentclass[a4paper,11pt]{article}
\pdfoutput=1 

\usepackage{jheppub}
\usepackage[T1]{fontenc} 
\usepackage{units}
\usepackage{numprint}
\usepackage{graphicx}
\usepackage{bm}
\usepackage[super]{nth}

\usepackage{lineno}
\title{\boldmath Neural-network-driven proton decay sensitivity in the $p\rightarrow \bar{\nu} K^{+}$ channel using large liquid argon time projection chambers}

 \author{C. Alt,}
 \author{B. Radics and}
 \author{A. Rubbia}
 \affiliation{Institute for Particle Physics and Astrophysics, ETH Z\"urich, CH-8093 Z\"urich, Switzerland}

\emailAdd{christoph.alt@cern.ch}

\abstract{We report on an updated sensitivity for proton decay via $p \rightarrow \bar{\nu} K^+ $ at large, dual phase liquid argon time projection chambers (LAr TPCs). Our work builds on a previous study in which several nucleon decay modes have been simulated and analyzed \cite{ABueno07}. At the time several assumptions were needed to be made on the detector and the backgrounds. Since then, the community has made progress in defining these, and the computing power available enables us to fully simulate and reconstruct large samples in order to perform a better estimate of the sensitivity to proton decay. In this work, we examine the benchmark channel $p\rightarrow \bar{\nu} K^{+}$, which was previously found to be one of the cleanest channels. Using an improved neutrino event generator and a fully simulated LAr TPC detector response combined with a dedicated neural network for kaon identification, we demonstrate that a lifetime sensitivity of $ \tau / \text{Br} \left( p \rightarrow \bar{\nu} K^+ \right) > \unit[7 \times 10^{34}]{years}$ at \unit[90]{\%} confidence level can be reached at an exposure of \unit[1]{megaton $\cdot$ year} in quasi-background-free conditions, confirming the superiority of the LAr TPC over other technologies to address the challenging proton decay modes.}

\begin{document} 
\maketitle

\flushbottom

\section{\label{sec:Intro}Introduction}
Direct experimental observation of proton decay would constitute evidence for Grand Unification, in which the electromagnetic, weak and strong interactions are combined into a single gauge with new massive bosons $X,Y$ as force carriers. The minimal SU(5) is the simplest Grand Unified Theory (GUT) and enables proton decay via the transformation of two up quarks into a lepton and anti-quark through the exchange of an $X$ boson, predicting a lifetime of $\tau / \text{Br} \approx \unit[10^{31}]{years}$ for $p \rightarrow  e^+ \pi^0$ \cite{HGeorgiSU51974}.\par
With the minimal SU(5) ruled out by recent results from Super-Kamiokande (SK) that constrain the partial proton lifetime to $\tau/ \text{Br} \left( p \rightarrow e^+ \pi^0 \right) > \unit[1.6 \cdot 10^{34}]{years}$ at 90 \% confidence level (CL) \cite{SuperKPositron2017}, supersymmetric extensions of Grand Unification (SUSY GUTs) gain more interest as they push the lifetime for $p \rightarrow e^+ \pi^0$ above the current experimental lower limit and open up new decay modes via the exchange of heavy supersymmetric particles \cite{GUTpe+pi0Ellis,SUSYGUTpe+pi0Arkani,GUTpe+pi0Hebecker,GUTpe+pi0Klebanov,SUSYGUTpK+nubarNath,SUSYGUTpK+nubarShafi,SUSYGUTpK+nubarLucas,SUSYGUTpK+nubarPati,SUSYGUTpK+nubarBabu,SUSYGUTpK+nubarAlciati}. The dominant decay mode in numerous SUSY GUTs is $p\rightarrow \bar{\nu} K^{+}$ with lifetimes of $10^{34} - 10^{35}$ years. Since the current best limit of $\tau/ \text{Br} \left( p \rightarrow \bar{\nu} K^+ \right) > \unit[5.9 \cdot 10^{33}]{years}$ at 90 \% CL by SK is below the predictions of SUSY GUTs \cite{SuperKPKaon2014}, searches for $p \rightarrow \bar{\nu} K^+$ remain of great interest.\par
Using a simplified simulation and making several assumptions on the detector design, we have found in a previous study that large dual phase (DP) LAr TPCs \cite{Rubbia:2004tz,Rubbia:2009md} can reach a lower lifetime limit of $\tau/ \text{Br}  > \unit[10^{35}]{years}$ at 90 \% CL in the $p \rightarrow \bar{\nu} K^+$ channel at an exposure of 1 megaton $\cdot$ year \cite{ABueno07}. In this paper, we update our result using an improved event generator, a well-defined design for a $\unit[{\sim}10]{kiloton}$ DP LAr TPC, a validated detector simulation based on data of the $\unit[3 \times 1 \times 1]{m^3}$ DP LAr TPC prototype \cite{311TechnicalPaper,ChristophsThesis}, a full reconstruction with aided pattern recognition and a neural-network-driven kaon identification. The DP LAr TPC detector design used in this study is considered as option for the Deep Underground Neutrino Experiment (DUNE) far detector complex, which will deploy a total of four $\unit[{\sim}10]{kiloton}$ single and dual phase LAr TPCs \cite{DUNETDRVol1}. The main improvements in the event generator are owed to more precise models for neutrino-nucleus interactions at the GeV scale that are tuned to recent high-statistics neutrino cross section measurements, see e.g. reference \cite{Alvarez-Ruso:2014bla}. In particular, the production of kaons in neutrino-nucleus interactions, which constitutes an important background for proton decay searches via $p \rightarrow \bar{\nu} K^+$, is better understood (see section \ref{sec:SignalBkg}).

%------------------------------------------------
\section{\label{sec:SimulationFramework}Simulation and reconstruction framework}

\subsection{\label{sec:SignalBkg}Signal and backgrounds}
Proton decay via $p \rightarrow \bar{\nu} K^+$ in argon constitutes the signal and atmospheric neutrino interactions with argon are considered as background. An accurate modeling of the argon nucleus is essential to the presented proton decay sensitivity study and the signal and background event samples are therefore simulated with the event generator toolkit GENIE \cite{GENIE}. The simulation workflows for both signal and background events include the modeling of the initial state of the argon nucleus in terms of nucleon density, momentum distribution and binding energy as well as the intranuclear propagation of particles emerging inside the nucleus. The momentum distribution and binding energy are modeled together with a so-called spectral function. Furthermore, the background simulation includes the atmospheric neutrino flux and neutrino-argon interaction models. Except for the neutrino flux, all aforementioned processes are implemented in GENIE and different models are available for each process. Consistent combinations of the interdependent processes are combined within so-called GENIE tunes, and the sensitivity study is carried out for signal and background samples generated with two different tunes in order to assess systematic uncertainties related to the event generation, see table \ref{TableGENIETunes1}.

\begin{table}[tbp]
\centering
\small
\begin{tabular}{ccc}
\hline 
GENIE tune & G18$\_$02a$\_$02$\_$11a  & G18$\_$10b$\_$00$\_$000 \\ 
 & (``reference tune'')  & (``alternative tune'') \\ 
\hline 
\hline
\multicolumn{3}{c}{$ $\vspace{-0.2cm}}  \\
& \hspace{1cm} \textbf{Signal $\&$ background} \hspace{1cm} & \\
\hline
Nucleon density distribution & Woods-Saxon \cite{PhysRev.95.577} & Woods-Saxon \\ 
\hline
Spectral function & GRFG BR \cite{PhysRevD.23.1070,Bodek:1981wr} & Local Fermi gas \\  
\hline
Intranuclear propagation & GENIE hA2018 & GENIE hN2018 \\ 
\hline
\multicolumn{3}{c}{$ $\vspace{-0.2cm}}  \\
& \textbf{Background} & \\
\hline
Atmospheric neutrino flux & HKKM2014 oscillated & HKKM2014 oscillated\\
\hline
\hline 
Elastic electron scattering & Marciano and Parsa \cite{Marciano:2003eq} & Marciano and Parsa \\ 
\hline 
Coherent scattering & Berger and Sehgal \cite{Berger:2008xs} & Berger and Sehgal \\ 
\hline 
Quasi-elastic scattering (NC) & Ahrens \cite{Ahrens:1986xe} & Ahrens \\ 
\hline 
Quasi-elastic scattering (CC) & Llewellyn-Smith \cite{LlewellynSmith:1971uhs} & Nieves \cite{Nieves:2004wx} \\ 
\hline 
Resonance production & Berger and Sehgal \cite{PhysRevD.76.113004} & Berger and Sehgal \\
\hline 
Meson exchange current (NC) & GENIE empirical & GENIE empirical \\ 
\hline 
Meson exchange current (CC) & GENIE empirical & Valencia \cite{PhysRevC.83.045501} \\ 
\hline 
Deep inelastic scattering & Paschos \cite{Paschos:2001np} & Paschos \\ 
\hline 
\end{tabular}
\normalsize
\caption{List of event generator models for the two GENIE tunes used in the proton decay sensitivity study. The abbreviation GRFG BR stands for global relativistic Fermi gas with Bodek-Ritchie extension, and CC and NC indicate charged current and neutral current neutrino interactions. The atmospheric neutrino flux simulation is not part of GENIE but mentioned in this table to provide a comprehensive overview of all models involved in the event generation. Most models in tune G18$\_$02a$\_$02$\_$11a are empirical, while the G18$\_$10b$\_$00$\_$000 tune uses more theoretically motivated models, making these two tunes a good combination to study event generator related uncertainties. Samples generated with the G18$\_$02a$\_$02$\_$11a tune are called reference samples in the following while those generated with the G18$\_$10b$\_$00$\_$000 tune are called alternative samples.}
\label{TableGENIETunes1}
\end{table}

The HKKM2014 atmospheric neutrino flux at solar maximum for the Sanford Underground Research Facility is used in both background samples \cite{Honda:2015fha}. The initial HKKM2014 flux is oscillated with the NuFit v4.1 neutrino oscillation parameters \cite{Esteban:2018azc}. The starting height of all neutrinos is set to \unit[15]{km} above the earth's surface and coherent forward scattering between neutrinos and electrons inside the earth is taken into account based on the earth density profile from the Preliminary Reference Earth Model (PREM) \cite{DZIEWONSKI1981297}.\par
For both signal samples, ${\sim} \numprint{100000}$ events are simulated and only the reference kaon decay mode $K^+ \rightarrow \mu^+ \nu_{\mu}$, which has a branching ratio of $\unit[63.6]{\%}$, is considered \cite{PDG}. The obtained results are assumed to be transferable to the remaining kaon decay modes, see section \ref{sec:Discussion}.
The reference background sample corresponds to an exposure of $\unit[10]{megaton \cdot years}$ and is used to tune the analysis cuts. The alternative background sample has a size of $\unit[2]{megaton \cdot years}$ and, together with the alternative signal sample, enables the determination of systematic uncertainties related to event generator models, which represent the dominant contribution to systematic uncertainties in this study. If not otherwise mentioned, only the reference signal and background samples are discussed in more detail in the following.

%\begin{figure}[tbp]
%\centering
%    \includegraphics[width=0.48\textwidth]{Figs/cAbsoluteKaonPlusKineticEnergyStatus1PaperAnswerToEditor2_G18_02a.png}
%    \hfill
%    \includegraphics[width=0.48\textwidth]{Figs/cCrossSectionTotalTimesFluxAllFlavorsPaper.png}
%    \caption{Left: signal $K^+$ kinetic energy distributions before and after intranuclear propagation (INP) for the reference tune. The vertical blue line at $\unit[105.3]{MeV}$ slackindicates the $K^+$ kinetic energy from the decay of a free proton at rest. Right: differential atmospheric neutrino energy spectrum for neutrino-argon interactions normalized to $\unit[1]{megaton \cdot year}$, resulting from the oscillated HKKM2014 flux tables and GENIE cross sections.}
%    \label{fig:KaonKineticEnergy}
%\end{figure}

\begin{figure}[tbp]
\centering
    \includegraphics[width=0.48\textwidth]{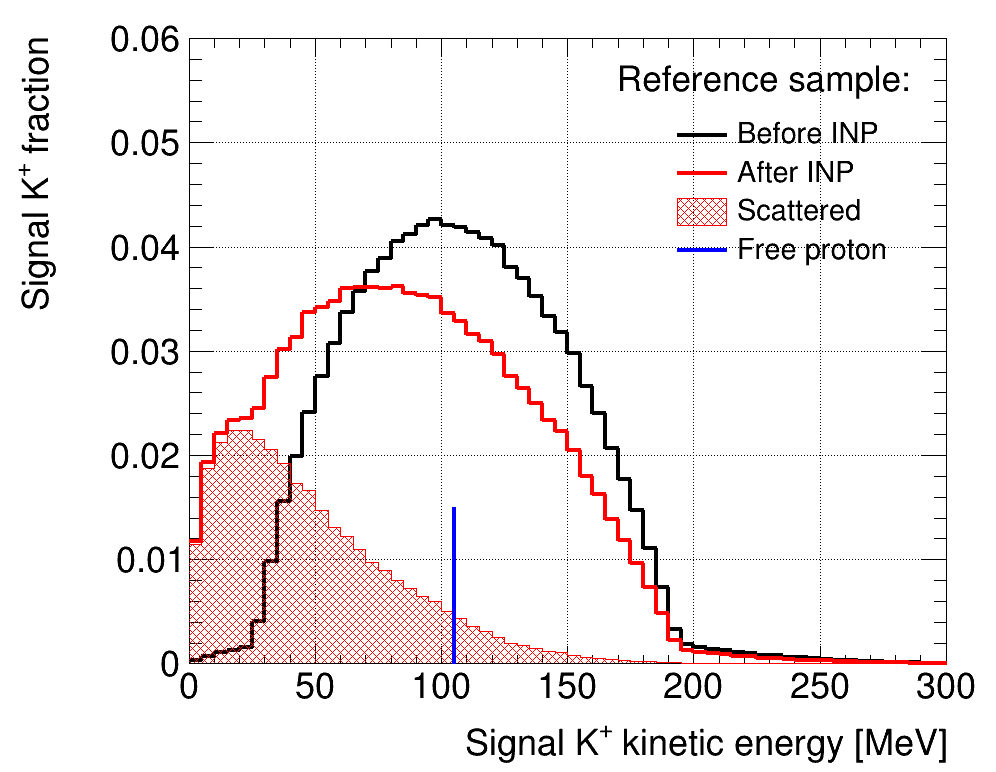}
    \hfill
    \includegraphics[width=0.48\textwidth]{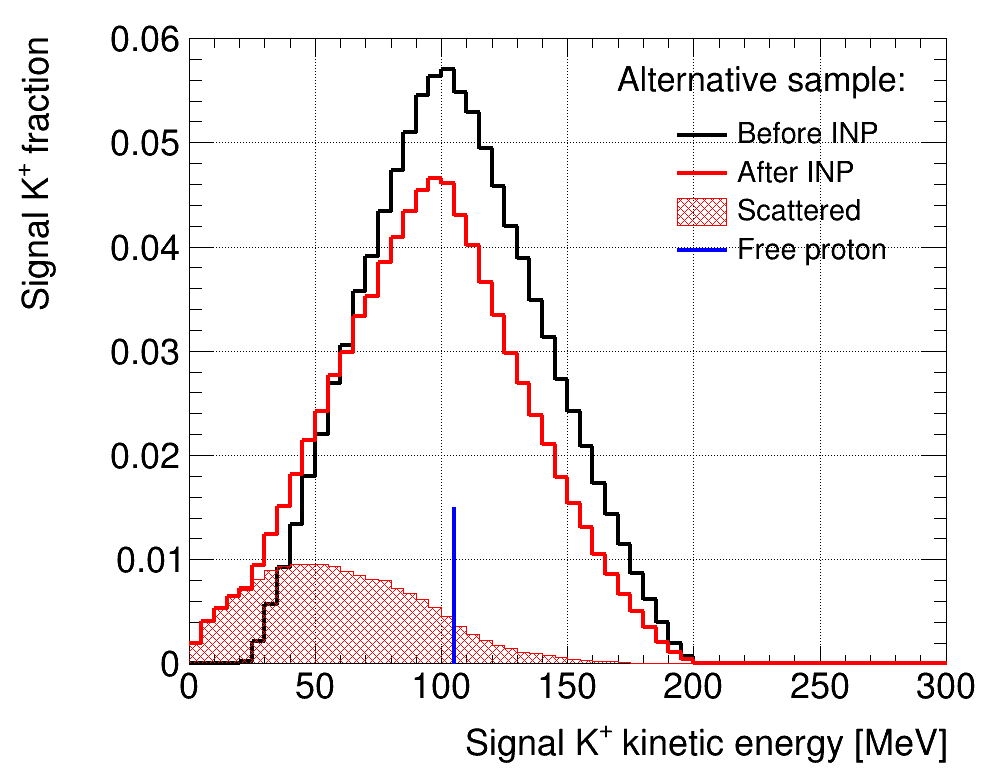}
    \caption{Signal $K^+$ kinetic energy distributions before and after intranuclear propagation (INP) for the reference (left) and alternative (right) samples. All distributions are normalized to their respective sample size before INP, and the simulation of charge-exchange in the alternative sample results in a signal $K^+$ loss of $\unit[15]{\%}$ after INP. The vertical blue line at $\unit[105.3]{MeV}$ indicates the $K^+$ kinetic energy from the decay of a free proton at rest.}
    \label{fig:KaonKineticEnergy}
\end{figure}

%The simulation of charge-exchange in the alternative sample results in a loss of $\unit[15]{\%}$ of $K^+$ and thus signal selection efficiency as the emerging $K^0$ are not attempted to be identified in the presented analysis, see section \ref{sec:Analysis}. On the other hand, the simulation of multi-nucleon knockouts in the reference sample shifts the distribution of scattered $K^+$ to lower energies with respect to the alternative sample as the energy loss of $K^+$ in multi-nucleon scatters is higher than in single-nucleon scatters. These counterbalancing effects lead to a higher signal selection efficiency in the alternative sample, see ... .

\begin{figure}[tbp]
\centering
    \includegraphics[width=0.8\textwidth]{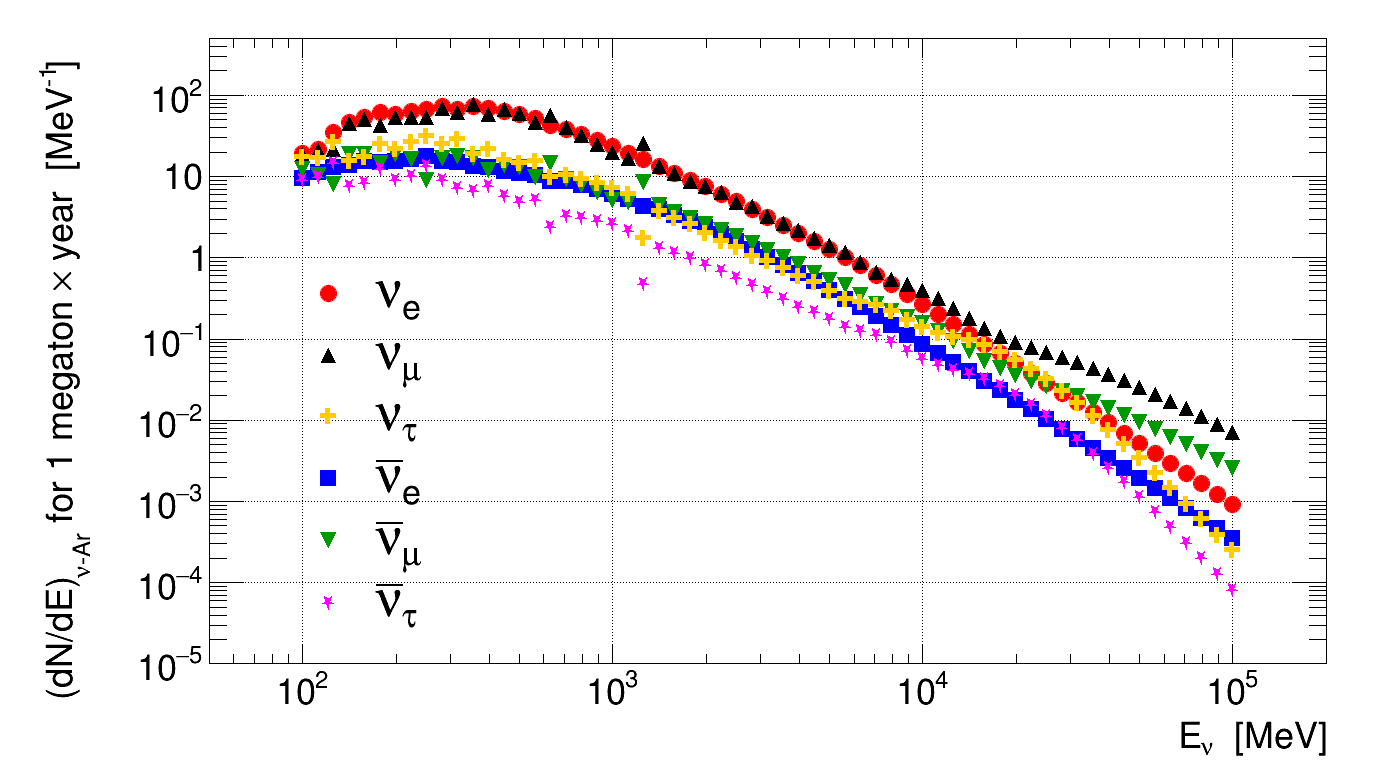}
    \caption{Differential atmospheric neutrino energy spectra for neutrino-argon interactions normalized to $\unit[1]{megaton \cdot year}$, resulting from the oscillated HKKM2014 flux tables and GENIE cross sections in the reference sample. The corresponding spectra in the alternative sample show only small deviations at the level of $\unit[1]{\%}$.}
    \label{fig:AtmNuInteractionRate}
\end{figure}

In both signal samples, the position of the decaying proton is sampled from the Woods-Saxon nucleon density distribution, and the $K^+$ is propagated through the nucleus in steps of $\unit[0.05]{fm}$. The interaction probability during each step is calculated with the local nucleon density and $K^+$-nucleon scattering cross sections that are obtained from fixed target kaon scattering experiments. No binding energy is subtracted from the proton at the time of the decay. As a result of empirical tuning inside GENIE, the binding energy $E_b = \unit[25]{MeV}$ is subtracted from the scattered kaon and nucleons, and added to the energy of the remnant nucleus, if the initial kinetic energy of the $K^+$ is greater than $\unit[100]{MeV}$. For scattered kaons with lower initial kinetic energy and for kaons that leave the nucleus without interaction, no binding energy is removed. Since the nucleon density and $K^+$ scattering cross sections are identical in both GENIE tunes, $\unit[32]{\%}$ of $K^+$ undergo a so-called final state interaction inside the remnant parent nucleus in both signal samples. The scattered $K^+$ typically lose a large amount of their kinetic energy to the struck nucleon, which makes their identification more difficult (see figure \ref{fig:KaonKineticEnergy}). In the hA2018 intranuclear propagation model used for the reference sample, $\unit[11]{\%}$ of all signal $K^+$ scatter off a single nucleon while $\unit[21]{\%}$ scatter off a multi-nucleon system. No charge-exchange is simulated and the scattered $K^+$ are therefore always present in the final state outside the nucleus, accompanied by low-energy neutrons and protons. The hN2018 model used for the alternative sample includes both elastic scatters off single nucleons and charge-exchange, with $\unit[17]{\%}$ of the signal $K^+$ undergoing elastic scatters and $\unit[15]{\%}$ charge-exchanging into a $K^0$ inside the nucleus. This results in an a priori signal selection efficiency loss of $\unit[15]{\%}$ in the alternative sample as the emerging $K^0$ is not attempted to be identified in the presented analysis, see section \ref{sec:Analysis}. On the other hand, the energy loss of $K^+$ in multi-nucleon scatters, which are only simulated in the reference sample, is higher than in single-nucleon scatters, resulting in a lower average $K^+$ kinetic energy after intranuclear propagation in the reference sample compared to the alternative sample (see figure \ref{fig:KaonKineticEnergy}). Since low-energy $K^+$ are more difficult to reconstruct and identify, the final signal selection efficiency in the reference sample is lower than in the alternative sample (see section \ref{sec:Sensitivity}).\par
The differential neutrino energy spectra for neutrino-argon interactions in the reference background sample, normalized to an exposure of \unit[1]{megaton $\cdot$ year}, are shown as a function of neutrino energy and neutrino flavor in figure \ref{fig:AtmNuInteractionRate}. The number of expected neutrino interactions is obtained by integrating the differential neutrino-argon interaction spectra, yielding a total of $\sim$\numprint{212000} interactions for \unit[1]{megaton $\cdot$ year}.\par
Six different charged current (CC) and neutral current (NC) neutrino-argon interactions are implemented in GENIE, see table \ref{TableGENIETunes1}. The by far most common produced particles in the various interactions are neutrons and protons, followed by pions, muons and electrons. The outgoing neutrinos in neutral current interactions are assumed to leave the detector without further interaction. Given the nature of the signal, the production of charged kaons is of special interest. One process through which charged kaons are produced is the so-called resonant associated kaon production. In a first step, the neutrino interacts with a nucleon as a whole to create a baryon resonance, a process important for neutrino energies between $\unit[1]{GeV}$ and $\unit[5]{GeV}$. In GENIE, the production amplitudes of 18 $N$ and $\Delta$ resonances with masses below $\unit[2]{GeV/c^2}$ are calculated with the Berger and Sehgal model. Relatively heavy resonances with masses $\gtrsim \unit[1.6]{GeV}$ can decay with a low probability into a $K^+$ or $K^0$ and an associated hyperon, typically a lambda ($\Lambda$) or sigma ($\Sigma$) baryon. The hyperons almost exclusively decay into a nucleon and a pion through the weak interaction \cite{PDG}. Both in the CC and NC resonant associated $K^+$ production, the hyperon and its decay products can be used to distinguish the interaction from the proton decay signal $p \rightarrow \bar{\nu} K^+$. For CC resonant $K^+$ production, an additional lepton is present. Resonant single kaon production without accompanying hyperons is possible in CC interactions if the exchanged $W^-$ boson turns an up quark into a strange quark to produce a strange baryon resonance that can decay into a neutral or negatively charged kaon and a nucleon. This process is not implemented in GENIE, but since it's Cabibbo suppressed and the charged lepton from the CC interaction makes it distinguishable from the signal, it is not expected to have a big impact on the presented results.\par
Deep inelastic scatters (DIS) can also give rise to associated and single kaon production, and both processes are implemented in GENIE. In DIS, the squared four-momentum transfer $Q^2$ is high enough for neutrinos to scatter off individual valence or sea quarks. The struck quark undergoes hadronization and typically produces several nucleons and pions. The radiated gluons involved in the hadronization process can produce strange-antistrange quark pairs that combine with spectator quarks to form kaons and associated hyperons. The hadronization process in GENIE is simulated with an empirical model for low energies and PYTHIA6 for high energies \cite{PYTHIA}. In CC DIS, an up quark can directly be transformed into a strange quark to produce neutral or negatively charged single kaons. Single kaons from DIS typically have higher energies than the $K^+$ from proton decay via $p \rightarrow \bar{\nu} K^+$ and the charged lepton produced in these interactions is another handle to distinguish them from proton decay.

\subsection{\label{sec:DetectorDesignAndSim}Detector design and simulation}
\begin{figure}[tbp]
    \centering
    \includegraphics[width=0.5\textwidth]{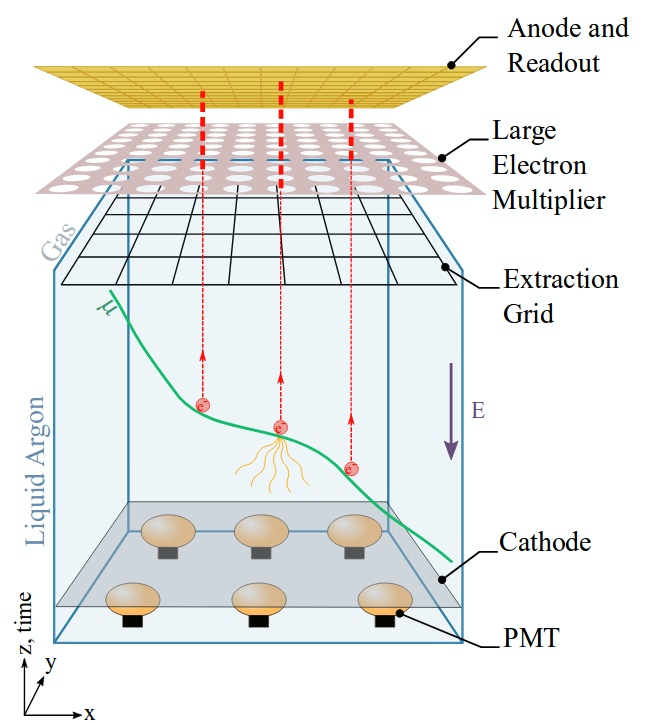}
    \caption{Sketch of a DP LAr TPC. The scintillation light produced by charged particles is collected by photomultiplier tubes (PMTs) at the bottom of the detector while the ionization charge is drifted upwards and amplified in argon gas before it is collected by two perpendicular readout views.}
    \label{fig:FigureDPSketch}
\end{figure}

The DP LAr TPC combines an active volume of liquid argon with a charge amplification and readout system in argon gas. Charged particles produce ionization charge and scintillation light as they travel through liquid argon. The ionization charge is drifted upwards and extracted into an argon gas layer by the means of electric fields. Inside the argon gas, the ionization charge is amplified inside so-called large electron multipliers and collected at the anode, see figure \ref{fig:FigureDPSketch}. 

The DP LAr TPC design used in this study has been defined in the context of an extensive R\&D program and is considered as far detector option for DUNE \cite{311TechnicalPaper}. The dimensions of the active volume are $\unit[60 \times 12 \times 12]{m^3}$ (length $\times$ width $\times$ height), providing an active mass of $\unit[{\sim}10]{kilotons}$ and an average of ${\sim}6$ atmospheric neutrino interactions per day. The $\unit[60 \times 12]{m^2}$ charge readout plane (CRP) consists of 80 independent $\unit[3 \times 3]{m^2}$ submodules, each surrounded by a gap of \unit[1]{cm}. Two perpendicular sets of readout channels with a pitch of \unit[3]{mm}, called view 0 and view 1, collect the charge signal in the submodules. The scintillation light is not considered in this study.\par
In order to reduce computation time, only nine CRP submodules, that are arranged as a square and yield a total charge readout area of $\unit[9 \times 9]{m^2}$, are considered in the simulation. The maximum drift distance remains \unit[12]{m}. The detector geometry is implemented in the LArSoft framework, a common LAr TPC software package for event simulation and reconstruction \cite{LArSoft}. The signal and background final state particle four-vectors obtained from GENIE are imported into LArSoft and placed \unit[6]{meter} below the center of the $\unit[9 \times 9]{m^2}$ CRP inside liquid argon at event time $t=0$. The energy loss, secondary interactions and decays of the final state particles are simulated in step sizes of \unit[$\sim$0.1]{mm} with GEANT4 \cite{GEANT4}. The local number of free electrons per unit length for each step is calculated from the step energy loss $dE$ and step length $ds$ with a modified version of Birks' law:

\begin{equation}
\frac{dN_e}{ds} = - \frac{dE}{ds} \cdot \frac{R}{W_e}
\label{eq:IonizationCharge}
\end{equation}

\noindent where $W_e = \unit[23.6]{eV}$ is the electron work function in liquid argon that equals to the average deposited energy necessary to produce one electron-ion pair \cite{MiyajimaWorkFunction} and $R$ the modified Birks' parameter that equals to the fraction of electron-ion pairs that do not recombine and therefore contribute to the charge signal:

\begin{equation}
R = \frac{A}{1 + \frac{1}{\rho} \frac{k}{\epsilon} \left( - \frac{dE}{ds} \right)}
\label{eq:ModifiedBirksLaw}
\end{equation}

\noindent with $\rho = \unit[1.4]{kg/l}$ the liquid argon density, $\epsilon = \unit[500]{V/cm}$ the nominal drift field and $- \left( dE / ds \right)$ the local linear stopping power. The parameter values $A=0.8$ and $k=\unit[0.0486]{kV \cdot MeV^{-1} \cdot g \cdot cm^{-3}}$ have been used \cite{ICARUSBirksParam}.\par
The electrons are drifted upwards from the center of each step. The drift time to the CRP is calculated with the drift velocity of \unit[1.6]{m/ms} at the nominal drift field of \unit[500]{V/cm} \cite{Agostino:2014qoa}. In order to account for longitudinal and transverse diffusion during the drift, the electron distribution at the CRP is smeared along the drift direction and in the plane perpendicular to the drift direction with a mean displacement $\lambda_\text{L,T} = \sqrt{2 \cdot D_\text{L,T} \cdot t_\text{Drift}}$, using the diffusion constants $D_\text{L} = \unit[0.62]{mm^2/ms}$ and $D_\text{T} = \unit[1.63]{mm^2/ms}$. The longitudinal diffusion constant $D_\text{L}$ has been measured by several experiments and the value used in this study is within the measured range \cite{Cennini:1994ha,Li:2015rqa,Agnes:2018hvf}. The transverse diffusion constant $D_\text{T}$ has been measured indirectly with high precision for drift fields above $\unit[2]{kV/cm}$. The extrapolation towards lower drift fields yields $D'_\text{T} \approx \unit[1.44]{mm^2/ms}$ at $\epsilon = \unit[500]{V/cm}$, which disagrees with the sparsely available data for low drift fields, and the used value of $D_\text{T} = \unit[1.63]{mm^2/ms}$ is thus a conservative estimate for the transverse diffusion \cite{Agostino:2014qoa}. The total gain in the CRP is set to 20 and the electrons are shared equally between the two readout views, with each electron being assigned to the closest readout channel in its respective readout view. All channels with at least one collected electron hold a waveform with the collected charge as a function of time. The charge waveform is shaped and transformed to a voltage waveform through convolution with the preamplifier shaping function $P_S(t)$:

\begin{equation}
    P_S(t) = \frac{P_G}{\tau_1 - \tau_2} \cdot \left( e^{-t/\tau_1} - e^{-t/\tau_2} \right)
\end{equation}

\noindent where $P_G = \unit[2.5]{mV/fC}$ is the preamplifier gain and $\tau_1 = \unit[2.83]{\mu s}$ and $\tau_2 = \unit[0.47]{\mu s}$ are the preamplifier shaping time constants. The voltage waveform is digitized in samples of \unit[400]{ns} with a \unit[12]{bit} ADC over a dynamic range of \unit[1800]{mV}. The preamplifier shaping function and gain are taken from pulsing measurements of the $\unit[3 \times 1 \times 1]{m^3}$ DP LAr TPC prototype at operating conditions \cite{311TechnicalPaper,ChristophsThesis}. Charge attenuation during the drift due to impurities and electronic noise are not simulated.\par
\begin{figure}[tbp]
\centering
    \includegraphics[width=0.48\textwidth]{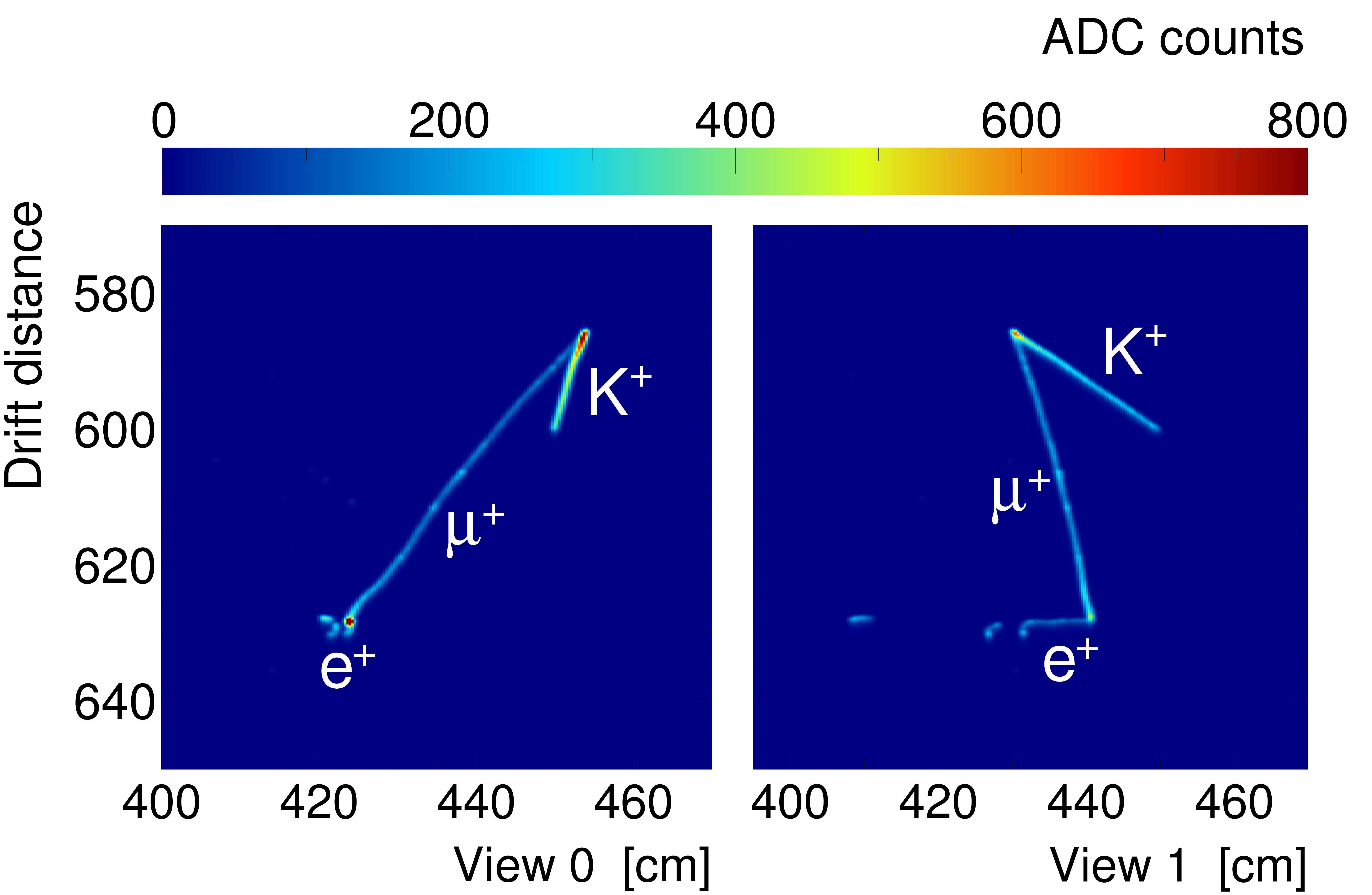}
    \hfill
    \includegraphics[width=0.48\textwidth]{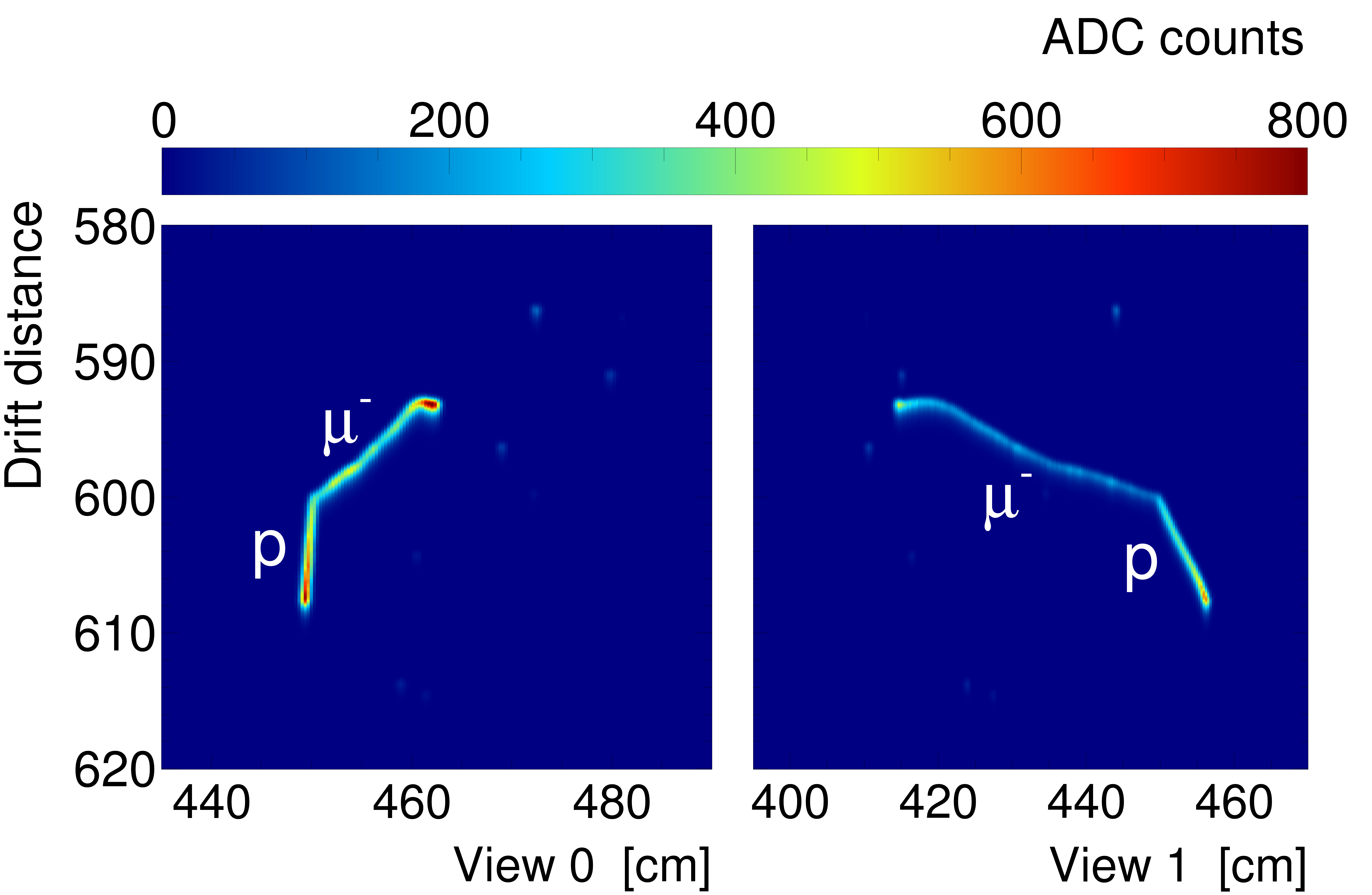}
    \caption{Example event displays for simulated proton decay via $p \rightarrow \bar{\nu} K^+$ (left) and for a $\nu_{\mu}$ charged current quasi elastic (CC QE) scatter on a neutron (right).}
    \label{fig:EventDisplay}
\end{figure}

Figure \ref{fig:EventDisplay} shows example event displays for proton decay via $p \rightarrow \bar{\nu} K^+$ with the kaon decaying into a $\mu^+$ and $\nu_{\mu}$ and for a $\nu_{\mu}$ CC quasi elastic (CC QE) scatter on a neutron, the most common background process. It will be shown in section \ref{sec:Analysis} that $\nu_{\mu}$ CC QE scatters on neutrons are an important background when the emerging proton is misidentified as signal $K^+$ and the muon has a similar energy as the $\mu^+$ from the $K^+$ decay. The event displays are a collection of ADC waveforms of neighboring channels, where the x-axes in both views directly correspond to the readout channel numbers and the drift distance on the y-axes is calculated by multiplying the drift time with the drift velocity. Thanks to the fine-grained imaging capability of LAr TPCs, all particles are clearly visible in the event display. The particle properties are reconstructed with the information stored in the waveforms and used in the analysis to distinguish proton decay from atmospheric neutrino background.

\subsection{\label{sec:Reco}Event reconstruction}
Hits are reconstructed by looking for peaks above threshold in the ADC waveforms. Peaks containing inflection points are split into separate hits. The hit charge $Q$ is determined by summing up all samples within a hit and is stored for further reconstruction.\par
The next step in the reconstruction is the identification of groups of hits that originate from the same particle. Reconstruction algorithms accomplish this task by looking for two types of patterns: continuous lines of hits originating from track-like particles such as kaons, protons, pions and muons, and discontinuous cone-like groups of hits from showering particles such as photons and electrons. As these pattern recognition algorithms for LAr TPCs are currently in development, an aided pattern recognition, in which hits originating from the same particle are grouped in both readout views with Monte Carlo truth information, is used in order to not limit the significance of this study by premature reconstruction algorithms. For all particles with at least two reconstructed hits in each of the two readout views, the 2D end points of the hit groups are matched between the readout views to obtain two 3D end points. The 3D track of the particle, hereafter simply called track, is defined as a straight line with length $L_\text{Track}$ that connects the two 3D end points. The end point in the half of the track with the lower charge content is defined as starting point of the track and the remaining end point as stopping point. The total charge of the track deposited in liquid argon $Q_\text{Track, LAr}$ is calculated in the readout view with most hits, also called best view, by summing up the charge of the contained hits and by correcting for the total gain in the CRP of 20 and the charge sharing between the two readout views. As both track-like and showering particles are reconstructed as straight tracks, the share of readout channels that do not contain hits associated to the track between its starting and stopping points $N_\text{Track, missing hits}$  is determined and later on used to distinguish track-like from showering particles.\par
The length of the track segment $ds$ from which a single channel has collected charge is calculated for the best view using the readout channel pitch and the direction of the track. The corresponding local charge deposition in liquid argon $dQ/ds$ and, through equations \ref{eq:IonizationCharge} and \ref{eq:ModifiedBirksLaw}, the local energy loss $-dE/ds$ are determined at each hit. The mean stopping power $\langle -dE/ds  \rangle$ and residual kinetic energy $E_\text{kin, residual}$ of the track at all hits are plotted against each other to obtain its stopping power profile, starting with the biggest hit near the track stopping point and walking along the trajectory towards the starting point by excluding the outermost hits with small charge content that originate from diffusing charge. The stopping power profiles are later on used for particle identification, see section \ref{sec:KaonID}.

\section{\label{sec:Analysis}Analysis}
The goal of this sensitivity study is to determine the lower proton lifetime limit per branching ratio $\tau / \text{Br} \left( p \rightarrow \bar{\nu} K^+ \right)$ for exposures up to \unit[1]{megaton $\cdot$ year} if no proton decay is observed. Since the lifetime limit typically increases with decreasing number of expected background events, a strong background rejection is essential to this study.\par
The analysis is carried out in three steps with the global strategy of identifying the signal $K^+$ and its decay products: event preselection, neural-network-driven track identification and final event selection.

\subsection{\label{sec:EventPreselection}Event preselection}
The event preselection uses reconstructed event variables to reject background events. The three following cuts are applied:

\begin{enumerate}
\item[1.1] Total number of hits in both views: $ 100 < N_\text{Event, Hits} < 800$
\item[1.2] Total charge in liquid argon of all hits in both views: $\unit[400]{fC} < Q_\text{Event, LAr} < \unit[\numprint{2000}]{fC}$ 
\item[1.3] Number of reconstructed tracks with $Q_\text{Track, LAr} > \unit[40]{fC}$ in the best view:\\ $ 3 \leq N_\text{Event, Tracks} \leq 4$
\end{enumerate}

\begin{figure}[tbp]
    \centering
    \includegraphics[width=0.59\textwidth]{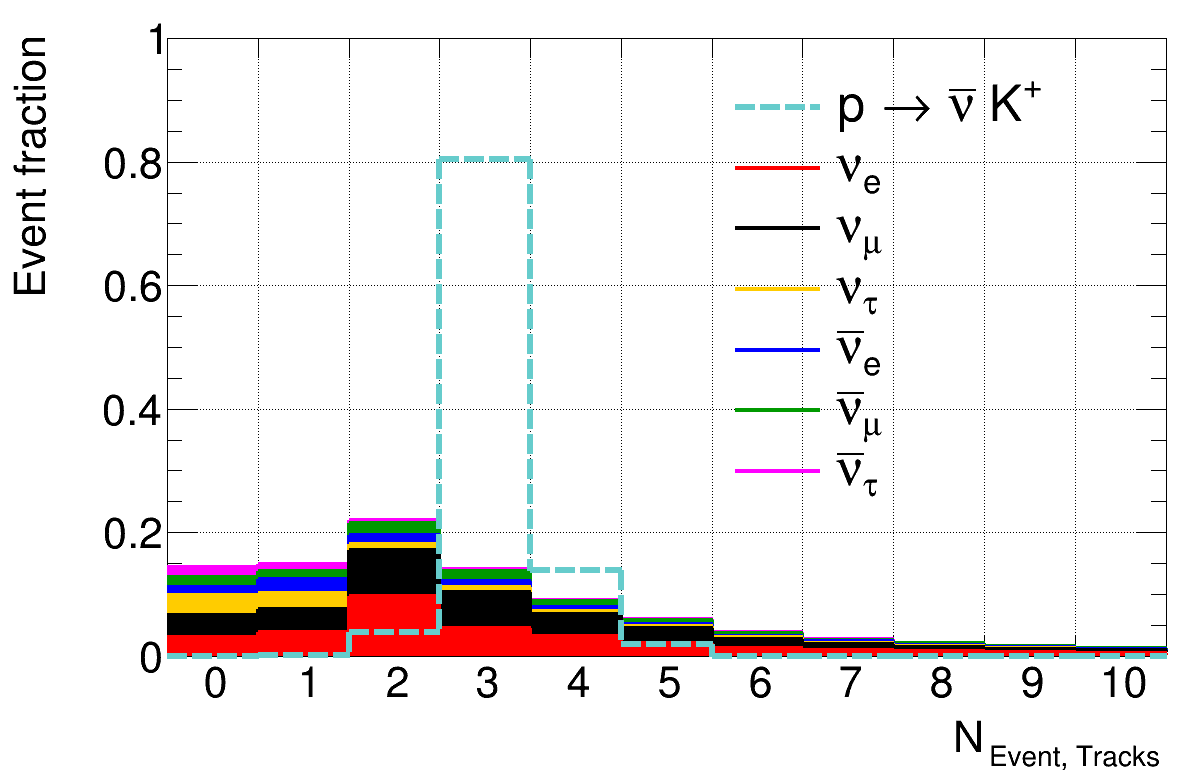}
    \caption{Multiplicity distribution of reconstructed tracks with $Q_\text{Track, LAr} > \unit[40]{fC}$ in the best view before event preselection in the reference signal and $\unit[10]{megaton \cdot years}$ background samples. The signal distribution reflects the kaon decay chain with a $K^+$, $\mu^+$ and $e^+$ as well as a potential proton knocked-out during the intranuclear propagation. The background distribution is dominated by quasi-elastic scatters which typically produce track multiplicities above threshold between zero and three, with the details depending on the neutrino flavor, scattered nucleon and intranuclear propagation.}
    \label{fig:EventPreselTrackMultiplicity}
\end{figure}

The ranges of cut 1.1 and 1.2 are chosen to include $\unit[99.9]{\%}$ of signal events while considerably reducing the number of background events. Cut 1.3 allows for three or four reconstructed tracks inside the event, which correspond to the signal $K^+$ and its daughter $\mu^+$ and $e^+$ as well as a potential proton knocked out during the intranuclear propagation. Only tracks with a reconstructed charge in liquid argon of $Q_\text{Track, LAr} > \unit[40]{fC}$ are considered in order to avoid low-energy photons that are emitted after neutron captures to dominate the track multiplicity distribution shown in figure \ref{fig:EventPreselTrackMultiplicity}. The signal selection efficiencies and total numbers of background events are shown in table \ref{tab:EventPresel} for the individual and combined event preselection cuts.\par

\begin{table}[tbp]
\vspace{0.1cm}
    \centering
    \begin{tabular}{c|ccc}
    \hline
    Cut \hspace{0.2cm} & \hspace{0.2cm} Signal selection efficiency \hspace{0.1cm} & \hspace{0.2cm} Background events & (efficiency) \\
    \hline
    \hline
    / & $\unit[100]{\%}$  & \numprint{2122620} & ($\unit[100]{\%}$)\\
    \hline
    \hline
    1.1 & $\unit[99.9]{\%}$ & \numprint{838806} & ($\unit[39.5]{\%}$)\\
    \hline
    1.2 & $\unit[99.9]{\%}$ & \numprint{674963} & ($\unit[31.8]{\%}$)\\
    \hline
    1.3 & $\unit[94.2]{\%}$ & \numprint{489663} & ($\unit[23.1]{\%}$)\\
    \hline
    \hline
    1 & $\unit[94.1]{\%}$ & \numprint{184365} & ($\unit[8.7]{\%}$)\\
    \hline
    \end{tabular}
    
    \caption{Signal and background selection efficiencies and total numbers of background events for event preselection cuts in the reference signal and $\unit[10]{megaton \cdot years}$ background samples. The cut labeled as 1 combines cuts 1.1, 1.2 and 1.3.}
    \label{tab:EventPresel}
\end{table}

\subsection{\label{sec:KaonID}Track identification}
The goal of the track identification is to determine the type of the particle that created a given track. Due to the nature of the signal and the consequent global analysis strategy of identifying the signal $K^+$ and its decay products, a simplified identification is used that only uses two classes of particles: signal $K^+$ vs. all other particles. In a first step, three preselection cuts are applied to tracks in events that survive the event preselection in order to select signal $K^+$-like tracks and reject tracks of all other particles. Subsequently, a neural network is used to determine how signal $K^+$-like the preselected tracks are. The track preselection variables and cuts are:

\begin{enumerate}
\item[2.1] Reconstructed track charge in liquid argon in the best view:\\ $\unit[40]{fC} < Q_\text{Track, LAr} < \unit[900]{fC}$
\item[2.2] Maximum share of readout channels without a hit assigned to the track between track starting and stopping point in both views: $N_\text{Track, missing hits} < \unit[1]{\%}$
\item[2.3] At least one hit in the best view that satisfies $\langle -dE/ds \rangle < \unit[20]{MeV/cm}$ and\\
$E_\text{kin, residual} < \unit[200]{MeV}$ in the stopping power profile
\end{enumerate}

The lower cut value for $Q_\text{Track, LAr}$ of $\unit[40]{fC}$ corresponds to a $K^+$ length of ${\sim} \unit[1]{cm}$ in liquid argon, which is the minimum track length for generating two hits in both readout views and thus for a successful particle identification. The upper cut value of $\unit[900]{fC}$ corresponds to the charge deposition of $K^+$ with the maximum kinetic energy $E_\text{kin} \approx \unit[200]{MeV}$, see figure \ref{fig:KaonKineticEnergy}.\par
Since $K^+$ are the highest ionizing particles in signal events, except for low-energy protons in the vertex region in events in which the $K^+$ underwent a final state interaction, the reconstructed signal $K^+$ track usually does not have missing hits from shadowing particles. Cut 2.2 is chosen accordingly to reject shower-like particles like electrons and photons which typically have a high share of missing hits.\par
Cut 2.3 defines a sensible range for the stopping power profiles used in the neural network classification. The upper limit in $\langle -dE/ds \rangle$ of $\unit[20]{MeV/cm}$ corresponds to the reconstructed stopping power of protons near their stopping point and $E_\text{kin, residual} = \unit[200]{MeV}$ is the maximum kinetic energy of signal $K^+$. The neural network will not attempt to classify tracks without at least one hit in this range. $\unit[76.4]{\%}$ of signal $K^+$ tracks survive the event and track preselection, which can be interpreted as signal selection efficiency at this stage of the analysis since every signal event contains exactly one $K^+$ in the reference sample.\par
\begin{figure}[tbp]
    \centering
    \includegraphics[width=0.48\textwidth]{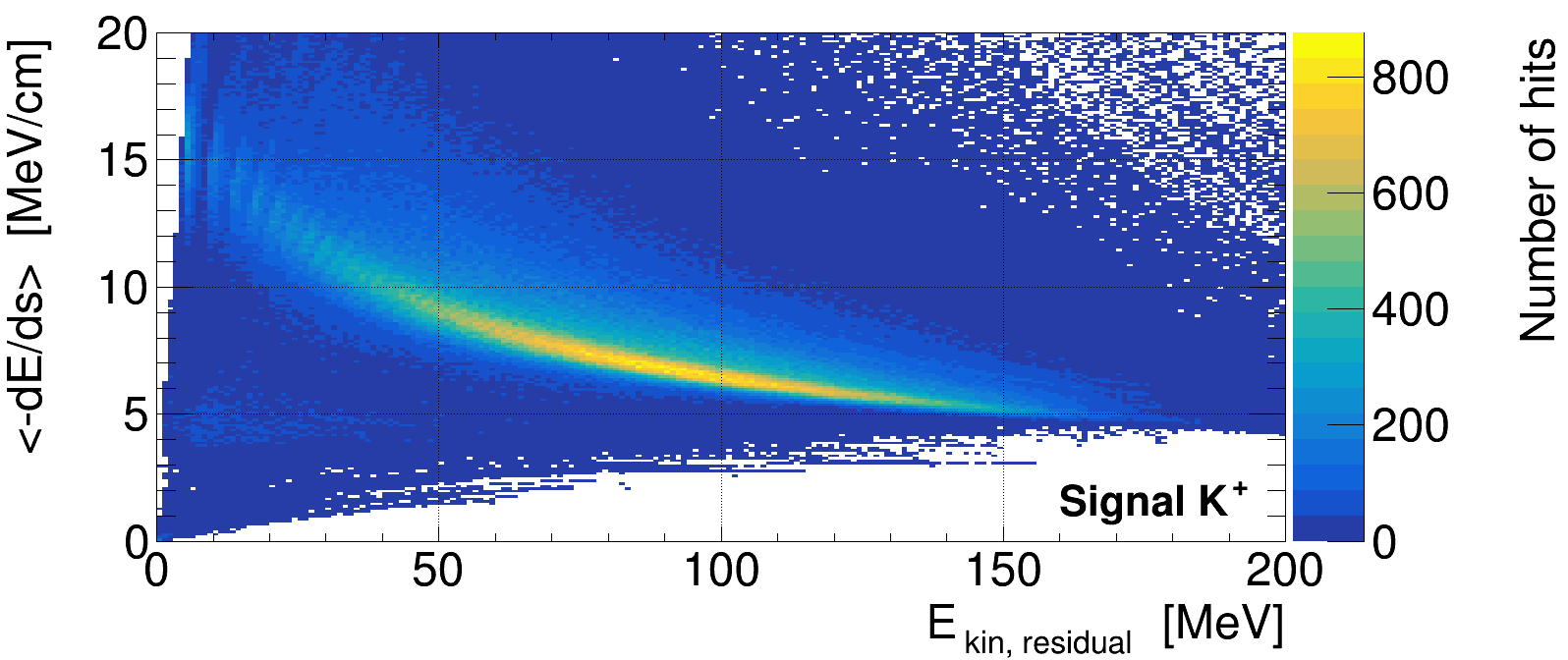}
    \hfill
    \includegraphics[width=0.48\textwidth]{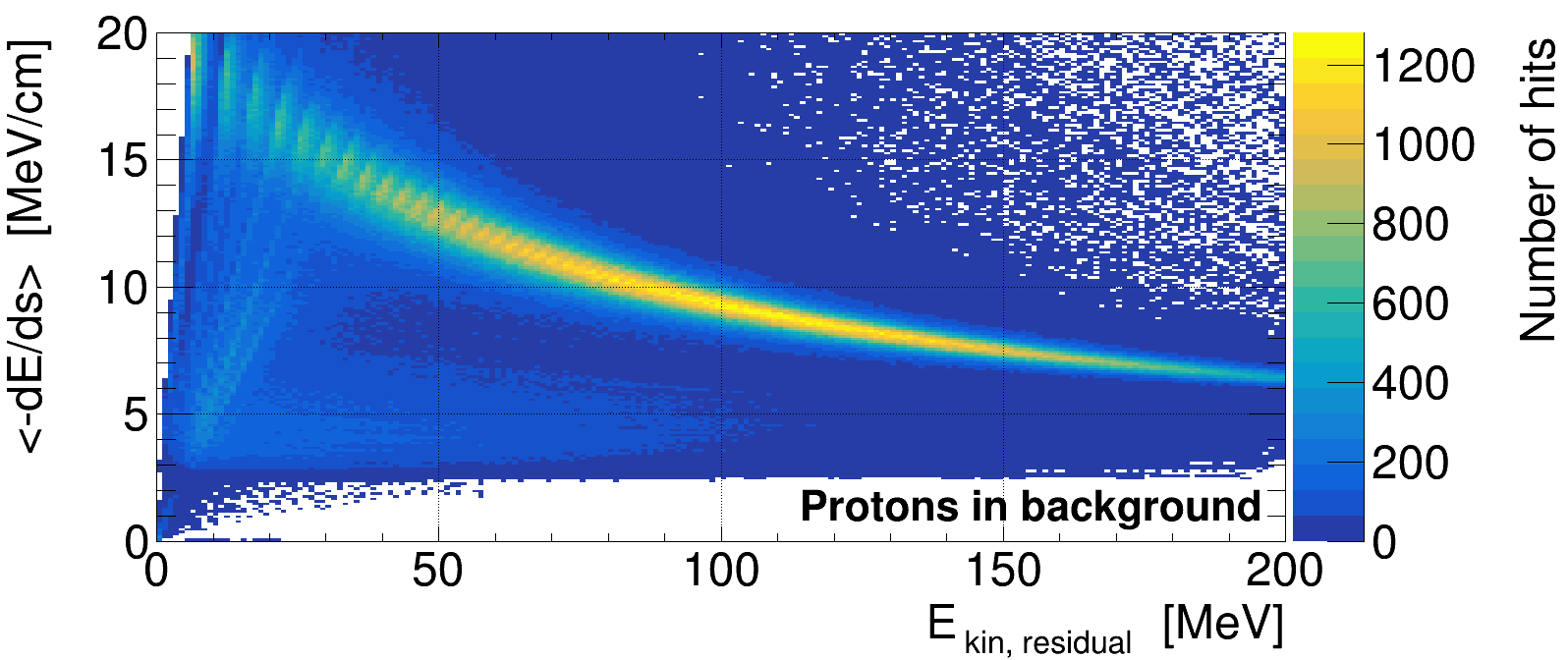}\\
    \includegraphics[width=0.48\textwidth]{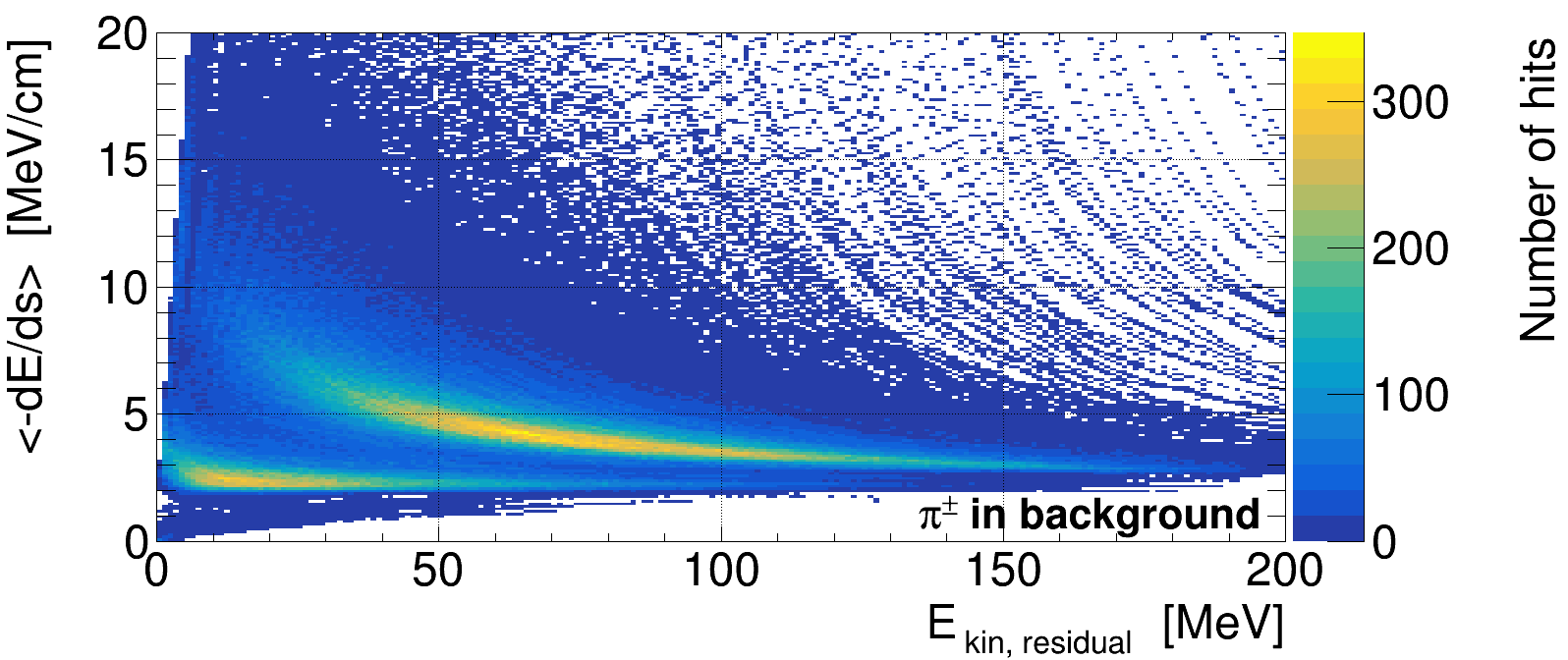}
    \hfill
    \includegraphics[width=0.48\textwidth]{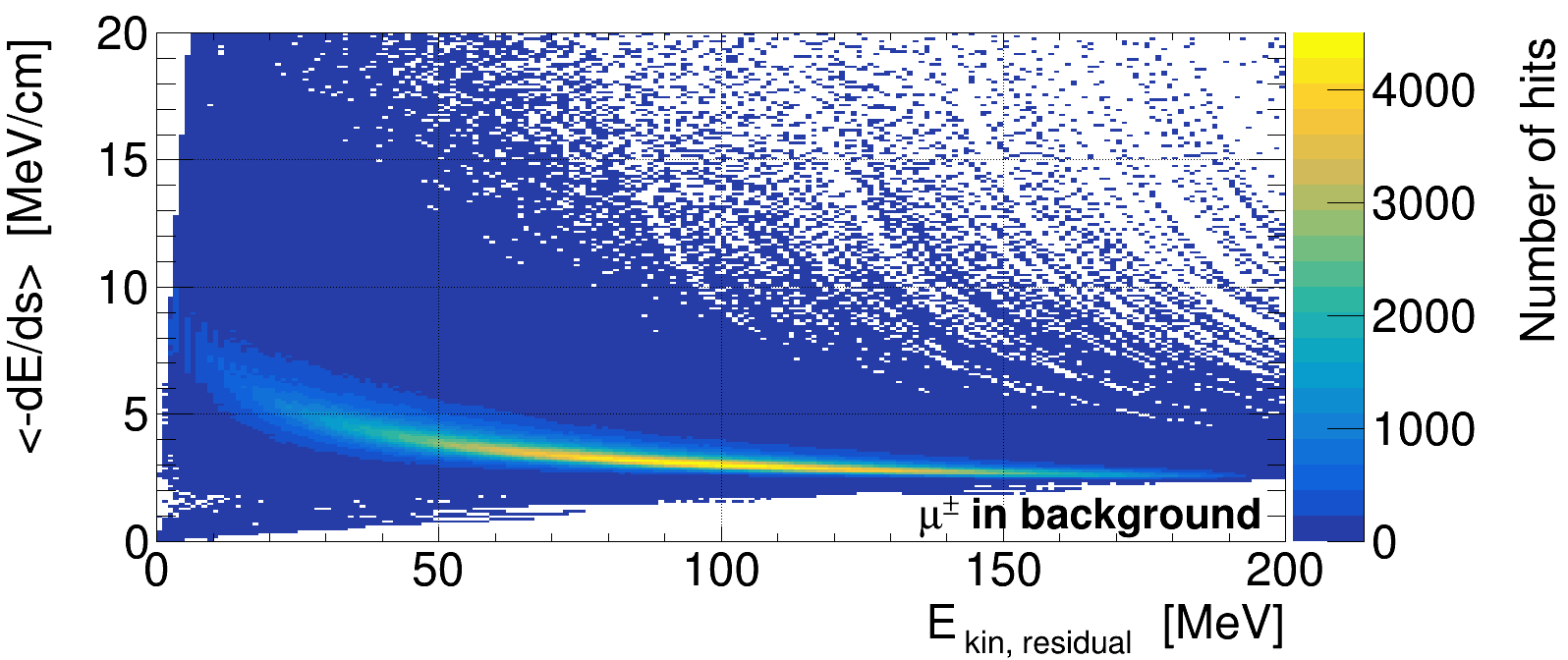}
    \caption{Track stopping power profiles with the mean stopping power $\langle -dE/ds \rangle$ and residual kinetic energy $E_\text{kin, residual}$ at each hit for $K^+$ in the reference signal sample (top left) and for protons (top right), pions (bottom left) and muons (bottom right) in the reference $\unit[10]{megaton \cdot years}$ background sample after event and particle preselection. In-flight decaying pions cause a secondary band at low stopping powers and residual kinetic energies without Bragg peak.}
    \label{fig:StoppingPowerProfiles}
\end{figure}

Tracks that pass the preselection are classified by a neural network that is trained with dedicated signal and background training samples generated with the reference GENIE tune. The training signal sample contains ${\sim}\numprint{65000}$ events while the training background sample corresponds to an exposure of $\unit[2]{megaton \cdot years}$. The neural network aims at distinguishing between signal $K^+$ and all other tracks in the signal and background samples. It is built using the TensorFlow library with an implementation of the Keras application programming interface \cite{Tensorflow,Keras}. The stopping power profiles of tracks that survive event and track preselection cuts are divided into 20 equally sized bins in $\langle -dE/ds \rangle$ between 0 and $\unit[20]{MeV/cm}$ and 20 equally sized bins in $E_\text{kin, residual}$ between 0 and $\unit[200]{MeV}$ to function as the 400-neuron input layer to the neural network. The stopping power profiles of signal $K^+$ as well as protons, pions and muons in the background sample that survive event and track preselection cuts are shown in figure \ref{fig:StoppingPowerProfiles}. The 400-neuron input layer is connected to the first inner layer with 64 neurons through the Rectified Linear Unit (ReLU) activation function, and the first inner layer is in turn connected through the ReLU activation function to the second inner layer, which also consists of 64 neurons. Finally, the second inner layer is connected to the output layer with 2 neurons through the softmax activation function. The two output neurons hold information about the signal $K^+$-likeness and signal $K^+$-unlikeness of a track. The softmax activation function forces the sum of both output values to 1 so that their information is redundant, and only the signal $K^+$-likeness output value is used for further analysis. The network is trained for a maximum of 40 epochs. During each epoch, $\unit[90]{\%}$ of the reshuffled training samples are used to train the network while the remaining $\unit[10]{\%}$ are used for validation. If the performance of the network on the validation subsample does not increase over 10 epochs, the training is complete.

\begin{figure}[tbp]
\centering
     \includegraphics[width=0.48\textwidth]{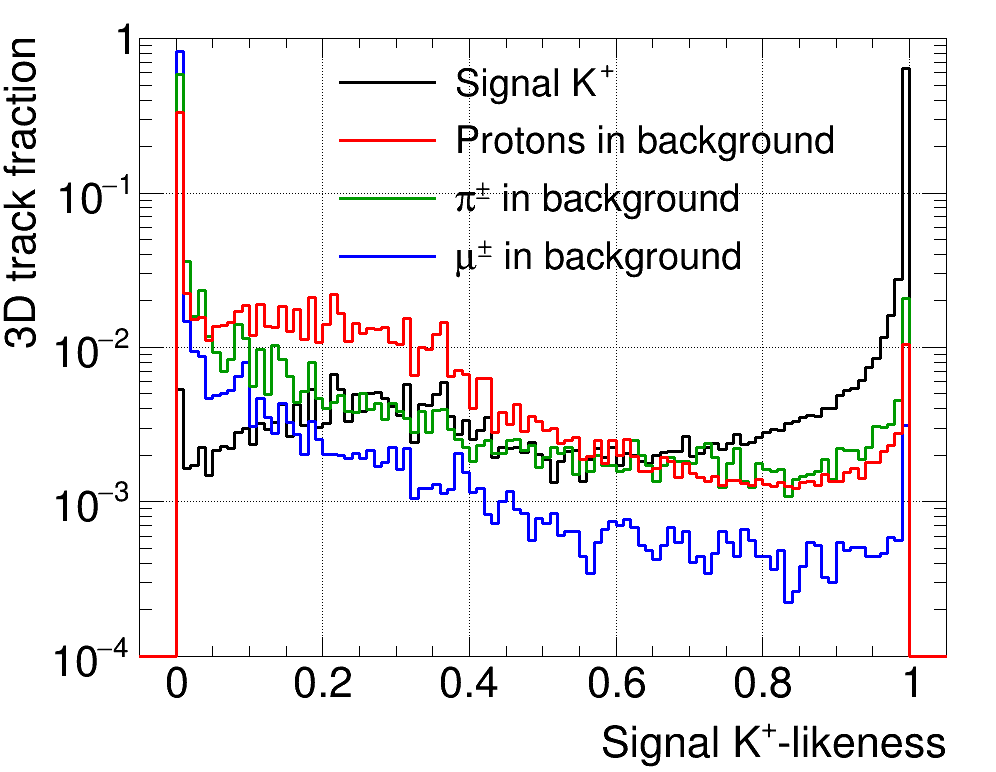}
     \hfill
     \includegraphics[width=0.48\textwidth]{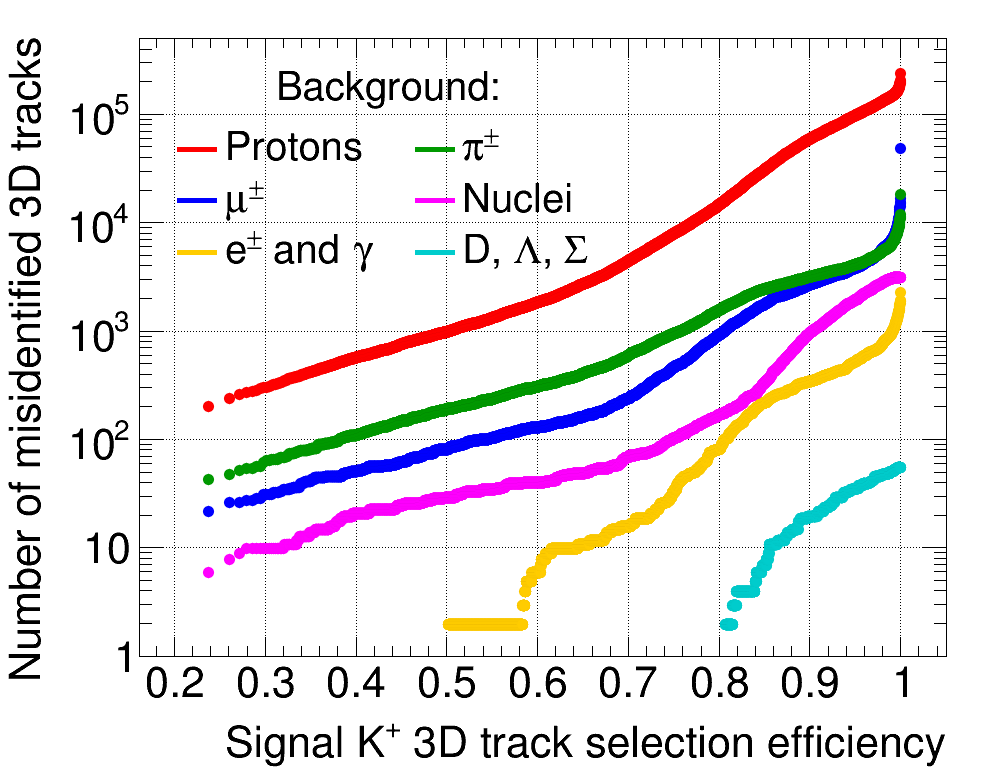}
    \caption{Left: Fraction of 3D tracks as a function of neural network signal $K^+$-likeness for $K^+$ in the signal sample and protons, pions and muons in the background sample. The inflection point at a signal $K^+$-likeness of 0.83 for protons, pions and muons in the background sample motivates the corresponding neural network cut in the final event selection. Right: number of 3D tracks in the background sample misidentified as signal $K^+$ as a function of signal $K^+$ track selection efficiency after event and particle preselection, using the signal $K^+$-likeness obtained from the neural network. Both figures show the reference signal and $\unit[10]{megaton \cdot years}$ background samples.}
    \label{fig:NNPerformance}
\end{figure}

After the training, the network is applied to all tracks that survive event and track preselection cuts in both the reference and alternative analysis samples. Only tracks with a signal $K^+$-likeness of 0.83 or higher are considered as signal $K^+$ in the final event selection in section \ref{sec:EventSel}, which represents the best compromise between signal $K^+$ track selection efficiency and rejection of other tracks (see left panel of figure \ref{fig:NNPerformance}). This cut value corresponds to a signal $K^+$ track selection efficiency of $\unit[75.9]{\%}$ for tracks that survive the event and track preselection, and thus to an overall signal $K^+$ track selection efficiency of $\unit[76.4]{\%} \cdot \unit[75.9]{\%} = \unit[58]{\%}$ after the neural network classification, with $\unit[76.4]{\%}$ being the efficiency after event and track preselection. The right panel of figure \ref{fig:NNPerformance} shows the signal $K^+$ track selection efficiency as a function of the number of tracks misidentified as signal $K^+$ in the background sample. Most misidentified tracks are protons since they are the most abundant charged particles in the background and the $K^+$ and proton stopping power profiles have a non-negligible overlap due to smearing effects from the detector simulation and reconstruction, see figure \ref{fig:StoppingPowerProfiles}.  At the cut value of \unit[75.9]{\%}, ${\sim}\numprint{10000}$ tracks are misidentified as signal $K^+$ in the full $\unit[10]{megaton \cdot years}$ reference background sample.

\subsection{\label{sec:EventSel}Final event selection}
In the final event selection, three cuts are applied to the preselected events that aim at the tracks from the signal $K^+$ and its daughter $\mu^+$ and $e^+$. A fourth cut only allows for potential low-energy protons from final state interactions in addition to the tracks from the $K^+$ decay chain. As opposed to cut 4.1, the targeted tracks in cuts 4.2 to 4.4 do not have to survive the track preselection and neural network classification. The four final event selection cuts are:

\begin{enumerate}
   \item[4.1] Exactly one track with a signal $K^+$-likeness of 0.83 or higher from the neural network classification, hereafter referred to as signal $K^+$ track.
  \item[4.2] Exactly one track that satisfies the following criteria, aiming at the $\mu^+$ from $K^+$ decay:
  \begin{enumerate}
    \item[4.2.1] \hspace{0.5cm} $\unit[520]{fC} < Q_\text{Track, LAr} < \unit[760]{fC}$
    \item[4.2.2] \hspace{0.5cm} $\unit[40]{cm} < L_\text{Track} < \unit[56]{cm}$
    \item[4.2.3] \hspace{0.5cm} $N_\text{Track, missing hits} < \unit[10]{\%}$
    \item[4.2.4] \hspace{0.5cm} Distance between track starting point and signal $K^+$ track stopping point:  $\text{ } \text{ } \text{ } \text{ } \text{ } D_{K^+} < \unit[5]{cm}$
    \item[4.2.5] \hspace{0.5cm} Angle to signal $K^+$ track in the best view: $ \alpha > \unit[10]{^{\circ}}$
  \end{enumerate}
  \item[4.3] Exactly one track that satisfies the following criteria, aiming at the $e^+$ from $\mu^+$ decay:
  \begin{enumerate}
    \item[4.3.1] \hspace{0.5cm} $N_\text{Track, missing hits} > \unit[10]{\%}$
    \item[4.3.2] \hspace{0.5cm} Number of hits in the best view: $N_\text{Track, Hits} > 10$
    \item[4.3.3] \hspace{0.5cm} $Q_\text{Track, LAr} > \unit[40]{fC}$
  \end{enumerate}
  \item[4.4] No additional track with:
  \begin{enumerate}
    \item[4.4.1] \hspace{0.5cm} $N_\text{Track, missing hits} < \unit[10]{\%}$
    \item[4.4.2] \hspace{0.5cm} $L_\text{Track} > \unit[5]{cm}$
    \item[4.4.3] \hspace{0.5cm} $Q_\text{Track, LAr} > \unit[40]{fC}$
  \end{enumerate}
\end{enumerate}
\vspace{0.2cm}

Cut 4.1 requires exactly one signal $K^+$-like track as it is the case in all signal events in the reference sample.\par
\begin{figure}[tbp]
    \vspace{0.35cm}
    \includegraphics[width=0.49\textwidth]{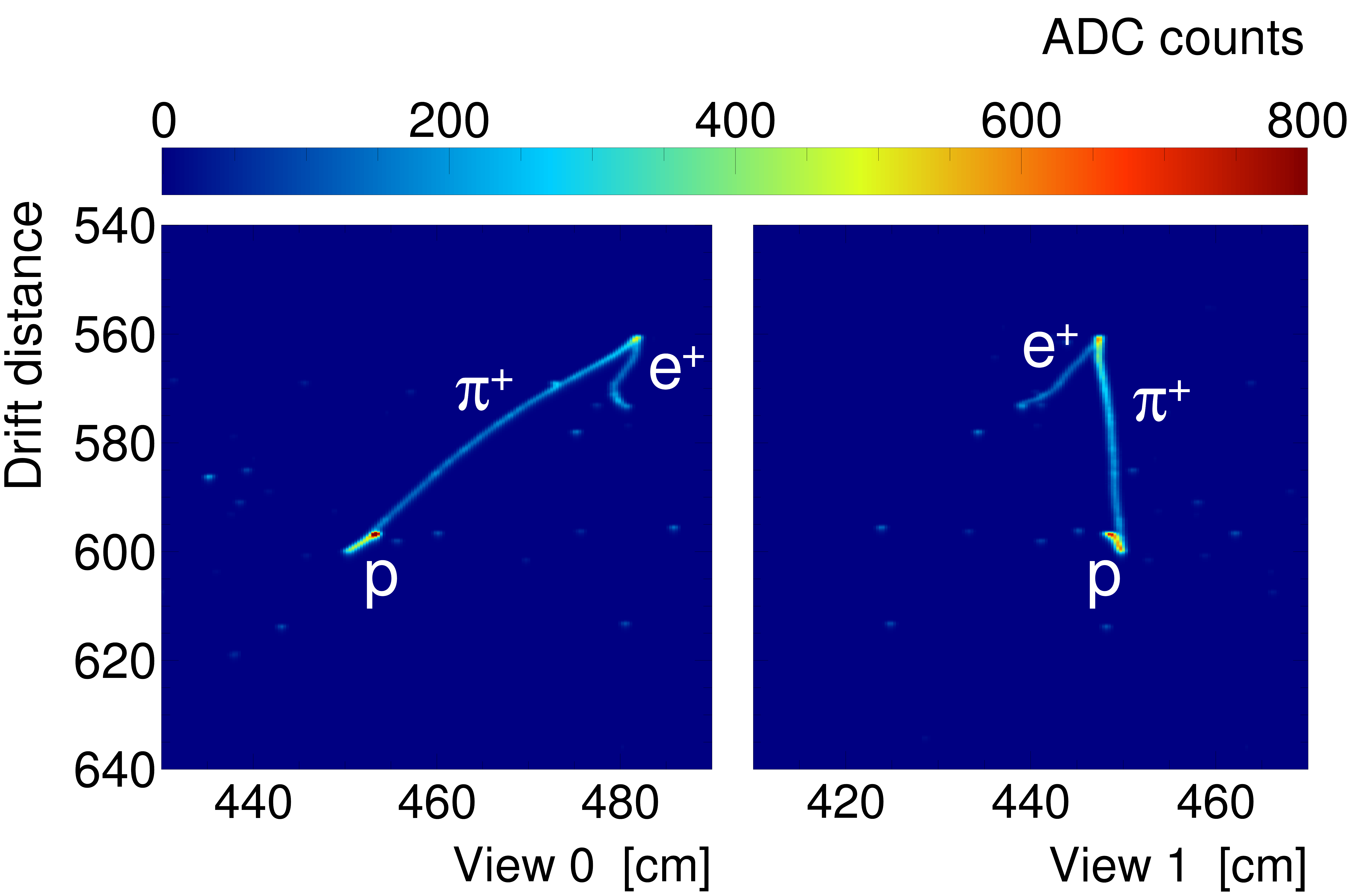}
    \hfill
    \includegraphics[width=0.49\textwidth]{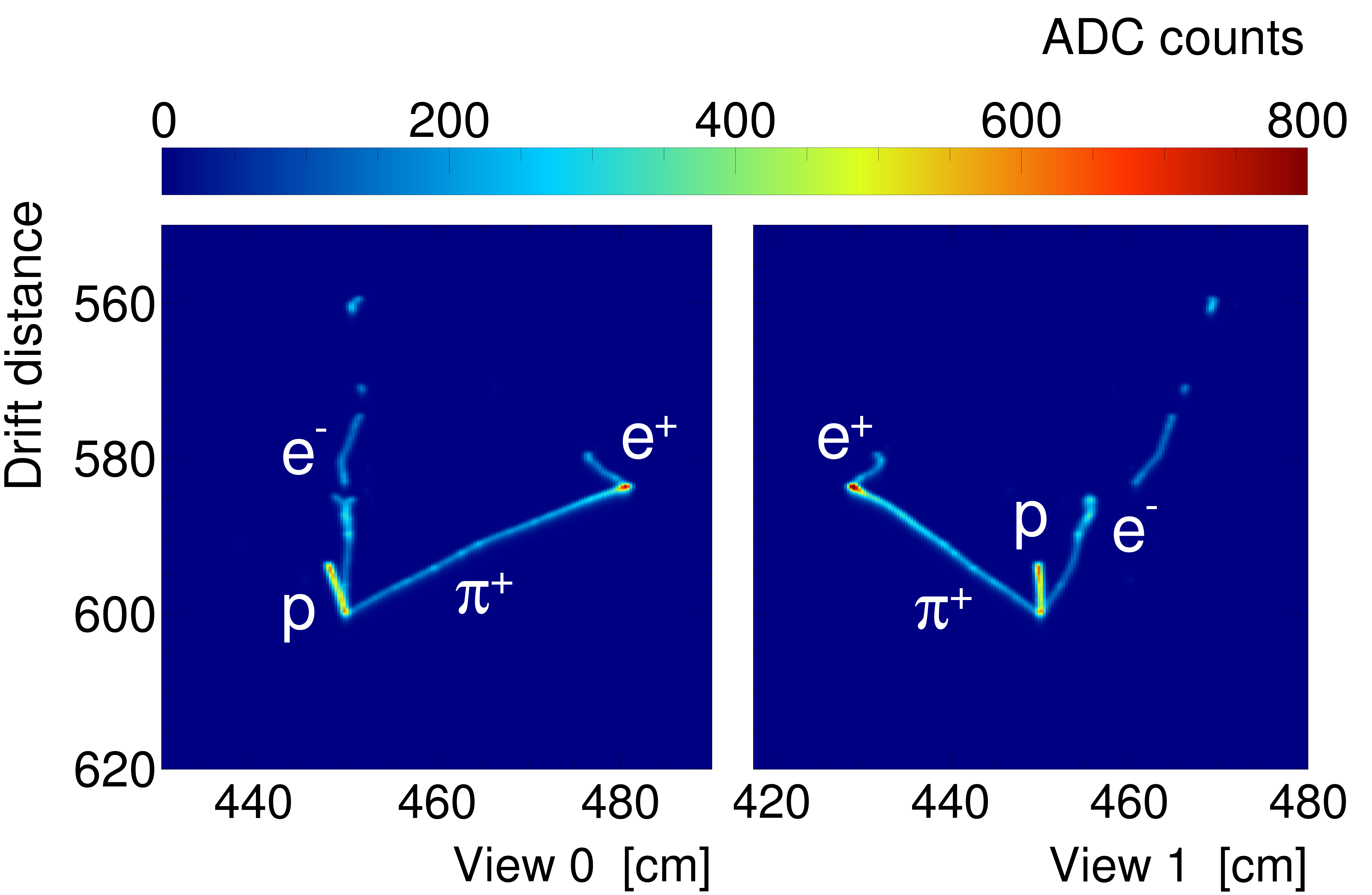}\\
    $ $\\
    \includegraphics[width=0.49\textwidth]{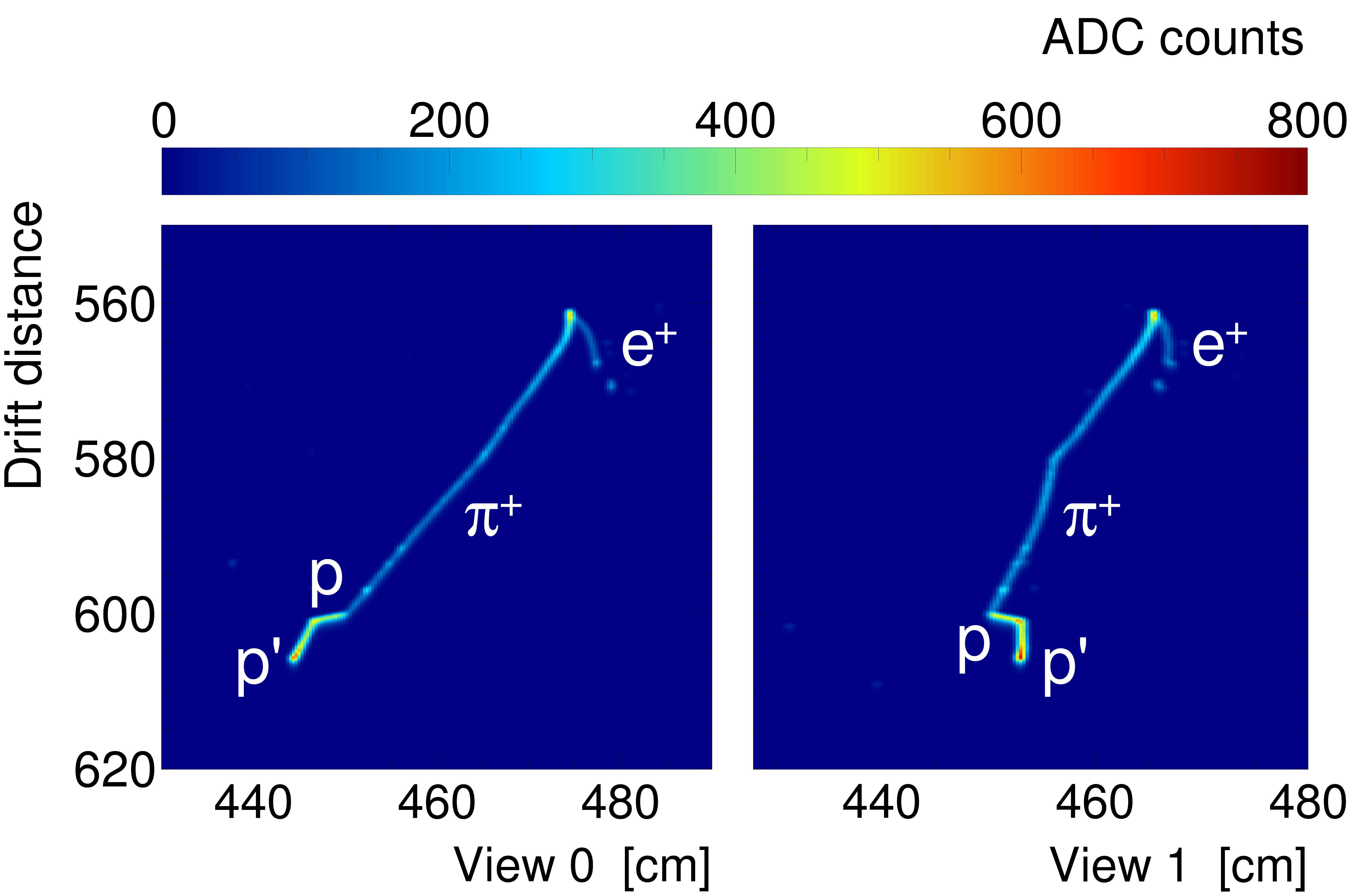}
    \hfill
    \includegraphics[width=0.49\textwidth]{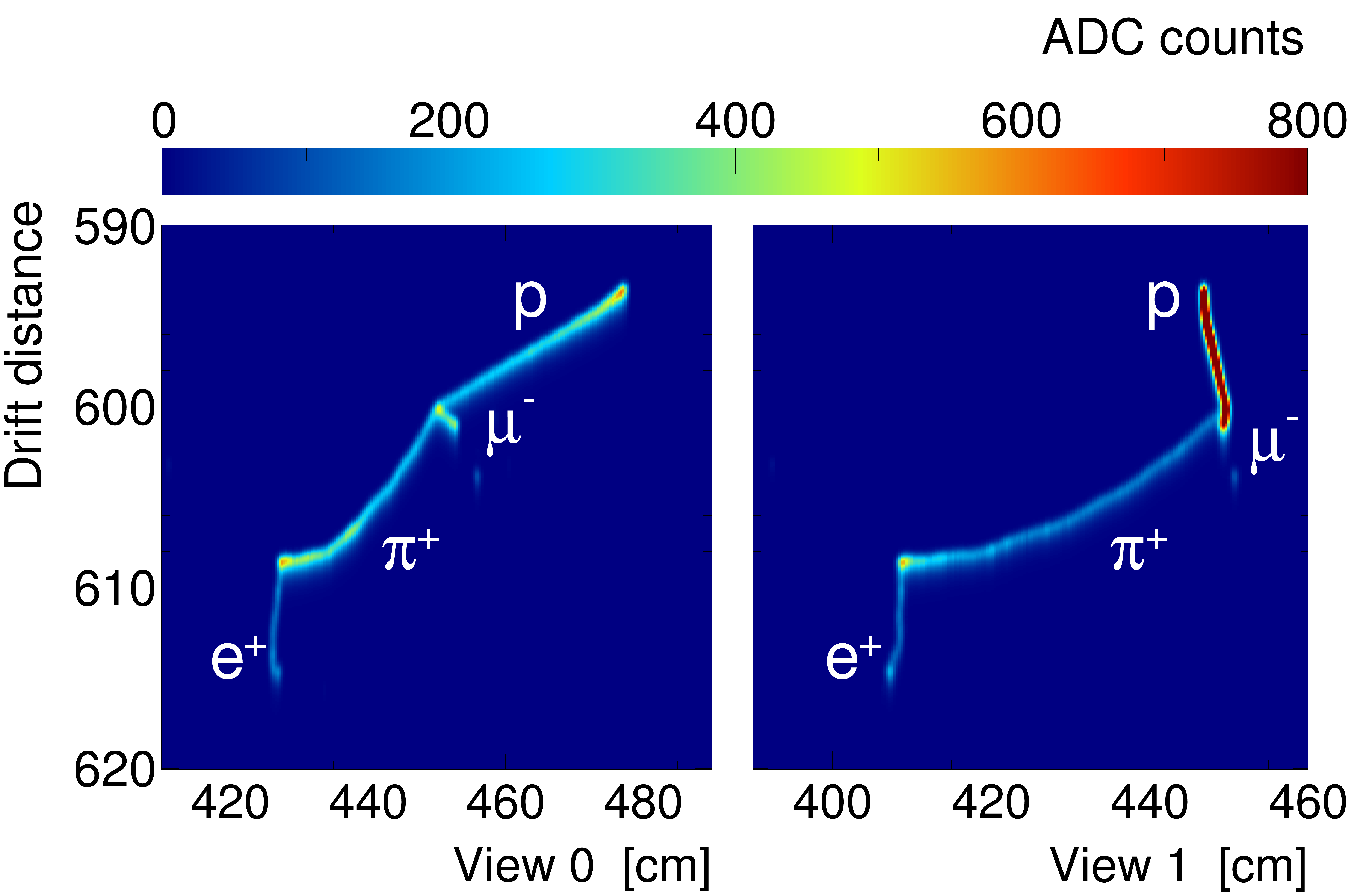}
    \caption{Event displays of persistent background events in the reference sample. In all events, the proton is misidentified as signal $K^+$ and the $\pi^+$ is mistaken for the $\mu^+$ from $K^+$ decay at rest. The top left event justifies cut 4.2.5 as the proton shadows the first part of the $\pi^+$ track in the best view. The top right event fails cut 4.3 as it has two showering particles and the two events at the bottom fail cut 4.4 since there is an additional track present, with the scattered proton in the bottom left event being reconstructed as two separate tracks.}
    \label{fig:EventDisplay2}
\end{figure}

Since $\unit[{\sim}92]{\%}$ of all signal $K^+$ decay at rest and only the two-body kaon decay mode $K^+ \rightarrow \mu^+ \nu_{\mu}$ is considered, cut 4.2 aims at monoenergetic $\mu^+$ with $E_\text{kin} = \unit[152.5]{MeV}$ and the corresponding charge deposition and length as chosen in cuts 4.2.1 and 4.2.2. As required by cuts 4.2.3 and 4.2.4, most $\mu^+$ tracks are track-like with less than $\unit[10]{\%}$ missing hits and close to the end point of the signal $K^+$ track. Cut 4.2.5 is introduced since some events in the background sample contain a proton that is misidentified as signal $K^+$ as well as a muon or pion with similar length and charge deposition as the $\mu^+$ from $K^+$ decay at rest, see top left event display in figure \ref{fig:EventDisplay2}. If the muon or pion travel in the same direction as the proton in these background events, the first part of their tracks are shadowed by the proton and it seems like the muon or pion emerge from the proton, just like the $\mu^+$ emerges from the $K^+$ decay. The minimum angle $\alpha$ under which two close tracks can be separated depends on the charge diffusion and the length of the track and was determined to $ \unit[10]{^{\circ}}$ for the values used in this study.

The Michel positron from the muon decay $\mu^+ \rightarrow e^+ \nu_{e} \bar{\nu}_{\mu}$ is typically reconstructed as a shower-like track with more than $\unit[10]{\%}$ missing hits, and cut 4.3.1 is set accordingly. In order to avoid low-energy photons in both signal and background samples to be misidentified as Michel positrons, cuts 4.3.2 on the charge and cut 4.3.3 on the number of hits in the best view are introduced.\par
Only low-energy proton tracks can be present in the signal sample in addition to the $K^+$, $\mu^+$ and $e^+$. Those tracks usually have less than $\unit[10]{\%}$ missing hits and are shorter than $\unit[5]{cm}$, and cuts 4.4.1 and 4.4.2 are chosen accordingly while cut 4.4.3 avoids low-energy photons. The signal selection efficiencies and number of background events after the consecutive final event selection cuts are shown in table \ref{TableFinalSelectionTune1}.

\begin{table}[tbp]
\vspace{0.8cm}
\centering
\begin{tabular}{c|ccc}
\hline
 Cut \hspace{0.2cm} & \hspace{0.2cm} Signal selection efficiency \hspace{0.1cm} & \hspace{0.2cm} Background events & (efficiency)\\
\hline
\hline
 / & $\unit[100]{\%}$ & \numprint{2122620} & ($\unit[100]{\%}$)\\
\hline
\hline
 1 & $\unit[94.1]{\%}$ & \numprint{184365} & ($\unit[8.7]{\%}$)\\
\hline
\hline
 4.1 & $\unit[58.0]{\%}$ & \numprint{9949} & ($\unit[0.5]{\%}$)\\
\hline
 & \multicolumn{3}{c}{$ $ \vspace{-0.5cm} }\\
 4.2 & $\unit[45.9]{\%}$ & \numprint{11} & $\left( 5 \cdot 10^{-6} \right)$ \\
  & \multicolumn{3}{c}{$ $ \vspace{-0.5cm} }\\
\hline
 & \multicolumn{3}{c}{$ $ \vspace{-0.5cm} }\\
 4.3 & $\unit[45.3]{\%}$ & \numprint{2} & $\left( 1 \cdot 10^{-6} \right)$ \\
  & \multicolumn{3}{c}{$ $ \vspace{-0.5cm} }\\
\hline
 & \multicolumn{3}{c}{$ $ \vspace{-0.5cm} }\\
 4.4 & $\unit[45.2]{\%}$ & \numprint{0} & $\left( < 5 \cdot 10^{-7} \right) $ \\
 \multicolumn{4}{c}{$ $ \vspace{-0.5cm} }\\
\hline
\end{tabular}
\caption{Signal and background selection efficiencies and number of background events for event preselection (cut 1) and consecutive final event selection cuts in the reference signal and $\unit[10]{megaton \cdot years}$ background samples. Although the background can be completely rejected in the studied sample, rare irreducible background events can not be excluded for larger exposures and the background efficiency is therefore given as upper limit after the last cut.}
\label{TableFinalSelectionTune1}
\end{table}

\begin{figure}[tbp]
    \centering
    \includegraphics[width=0.48\textwidth]{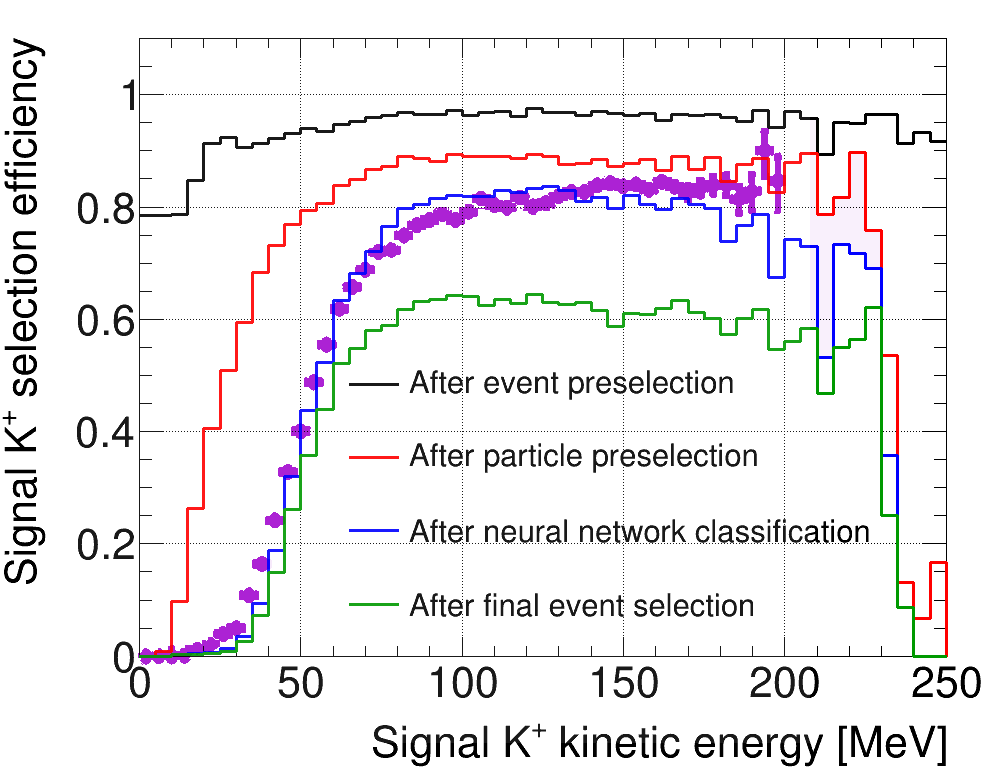}
    \hfill
    \includegraphics[width=0.48\textwidth]{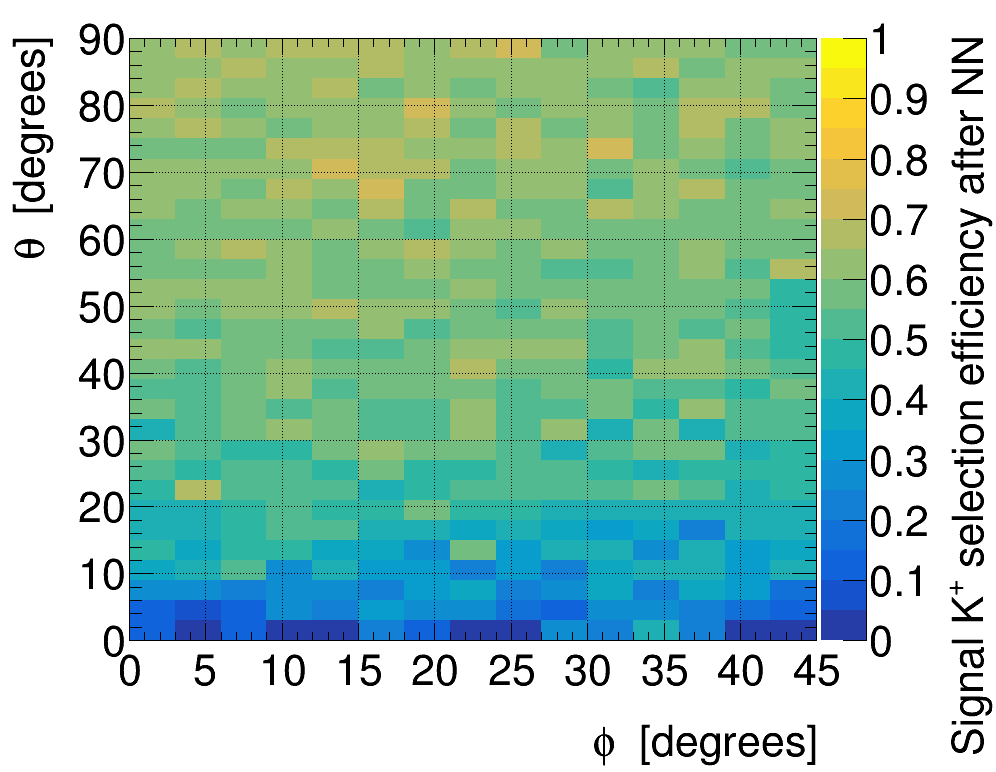}
    \caption{Left: signal $K^+$ selection efficiency as a function of true kinetic energy throughout the analysis. Since every signal event contains exactly one $K^+$ in the reference sample, the y-axis can also be interpreted as signal selection efficiency. The purple points show the signal $K^+$ tracking efficiency for a similar study reported in \cite{Abi:2020kei} and are put into context at the end of section \ref{sec:Sensitivity}. Right: signal $K^+$ selection efficiency as a function of true $K^+$ start direction after the neural network (NN) classification. The ranges of $\theta$ and $\phi$ have been downsized by exploiting different symmetries in the detector: $\phi = \unit[0]{^{\circ}}$ is parallel to the readout strips in one of the readout views and $\phi = \unit[45]{^{\circ}}$ is in the middle of both readout view orientations. $\theta = \unit[0]{^{\circ}}$ is parallel and antiparallel to the drift direction and $\theta = \unit[90]{^{\circ}}$ is parallel to the charge readout plane. The selection efficiency decreases significantly for kaons that travel parallel or antiparallel to the drift direction ($\theta = \unit[0]{^{\circ}}$) but is stable for kaons that travel parallel to the readout strips in one of the readout views ($\phi = \unit[0]{^{\circ}}$).}
    \label{fig:SignalKSelEff}
\end{figure}

The losses in signal selection efficiency are mainly due to low-energy $K^+$ that have scattered inside the nucleus, badly reconstructed $K^+$ traveling parallel or antiparallel to the drift direction and in-flight decaying $K^+$. Figure \ref{fig:SignalKSelEff} shows the signal $K^+$ selection efficiency as a function of the true $K^+$ kinetic energy throughout the analysis and as a function of true kaon direction after the neural network classification. The selection efficiency for low-energy $K^+$ drops significantly after particle preselection, which can be explained by cut 2.1 that requires a minimum charge deposition in liquid argon of $\unit[40]{fC}$ per track as well as by cut 2.3 that rejects kaon tracks with unreasonable stopping power profiles, which is more likely to occur for short tracks from low-energy $K^+$. A similar drop is observed after the neural network classification since the direction of short tracks is more likely to be misreconstructed, which leads to shifted stopping power profiles. These effects are enhanced by the diffusion of the drifting charge and could be mitigated by a better spatial resolution, see section \ref{sec:DetectorDesignAndSim}. Additional inefficiencies are introduced by $\mu^+$ traveling in the same direction as the parent $K^+$ (cut 4.2.5) and low-energy Michel positrons from the $\mu^+$ three-body decay that are not reconstructed and lead to a signal track multiplicity of 2 (cut 1.3 in the event preselection).

In ten out of the eleven background events that pass cut 4.2, a proton is misidentified as signal $K^+$ and in nine events, a charged pion is mistaken for the $\mu^+$ from the $K^+$ decay. Out of the eleven events, six have no shower-like tracks as defined by cut 4.3 since they contain only negatively charged pions or muons and the $\mu^-$ is captured by an argon atom without producing a Michel electron, and three events have two shower-like tracks instead of one. The remaining two events with one shower-like track that pass cut 4.3 have an additional track that fails cut 4.4. Figure \ref{fig:EventDisplay2} shows four event displays of persistent background events in which the proton is misidentified as signal $K^+$ and the $\pi^+$ is mistaken for the $\mu^+$ from $K^+$ decay at rest. Even the most persistent background events are clearly distinguishable from proton decay via $p \rightarrow \bar{\nu} K^+$ in the event display since the misidentified proton shares the same vertex with other particles and its Bragg peak is not connected to a second track, as it is the case for the $K^+$ in the signal sample.\par
The same analysis from event preselection to final event selection is applied to the alternative signal and background samples (see table \ref{TableGENIETunes1}), yielding a signal selection efficiency of $\unit[46.8]{\%}$ and 0 background events in $\unit[2]{megaton \cdot years}$.

\section{\label{sec:Sensitivity}Proton decay sensitivity results}
The lower lifetime limit per branching ratio for $p \rightarrow \bar{\nu} K^+$ can be obtained with:

\begin{equation}
    \tau / \text{Br} \left( p \rightarrow \bar{\nu} K^+ \right) > T \cdot N_p \cdot \epsilon \cdot \frac{1}{S} \hspace{0.2cm}
    \label{eq:sensitivity}
\end{equation}

\noindent where $T$ is the exposure in kiloton $\cdot$ years, $N_p = 2.7 \cdot 10^{32}$ the number of protons in one kiloton of argon, $\epsilon$ the signal selection efficiency and $S$ the upper limit on the number of signal events at $\unit[90]{\%}$ confidence level (CL) that depends on the number of observed events $N$ and the number of expected background events $B$. In the previous section it has been shown that the background can be reduced to 0 for both samples. Since the considered exposures of $\unit[10]{megaton \cdot years}$ and $\unit[2]{megaton \cdot years}$ are beyond the expectation for DUNE, the proton decay sensitivity is only calculated for exposures up to $\unit[1]{megaton \cdot year}$, which is a conservative estimate for the maximum achievable exposure with DUNE. The neural network cut 4.1 in the final event selection is adjusted to obtain $B=0.5$ background events at exposure steps of $\unit[200]{kiloton \cdot years}$ for both samples separately, and the concomitant signal selection efficiencies are summarized in table \ref{TableSensitivitySignalEff}.

\begin{table}[tbp]
\centering
\begin{tabular}{ccc}
\hline
\hspace{0.3cm} Exposure \hspace{0.6cm} & \multicolumn{2}{c}{Signal selection efficiency} \\
\multicolumn{3}{c}{$ $ \vspace{-0.4cm} }\\
 & \hspace{0.35cm} reference sample & \hspace{0.2cm} alternative sample\\
\hline
\hline
 $\unit[200]{kiloton \cdot years}$  & $\unit[51.9]{\%}$ & $\unit[53.6]{\%}$ \\
 \hline
  $\unit[400]{kiloton \cdot years}$  & $\unit[50.7]{\%}$ & $\unit[53.3]{\%}$\\
 \hline
  $\unit[600]{kiloton \cdot years}$  & $\unit[49.9]{\%}$ & $\unit[53.2]{\%}$\\
 \hline
  $\unit[800]{kiloton \cdot years}$  & $\unit[49.6]{\%}$ &$\unit[52.9]{\%}$ \\
 \hline
 $\unit[1]{megaton \cdot year}$  & $\unit[49.1]{\%}$ & $\unit[52.9]{\%}$ \\
\hline
\hline
\end{tabular}
\caption{Signal selection efficiencies in the reference and alternative samples with neural network cut 4.1 of the final event selection tuned for $B=0.5$ background events at selected exposures.}
\label{TableSensitivitySignalEff}
\end{table}

\begin{figure}[tbp]
    \centering
    \includegraphics[width=0.6\textwidth]{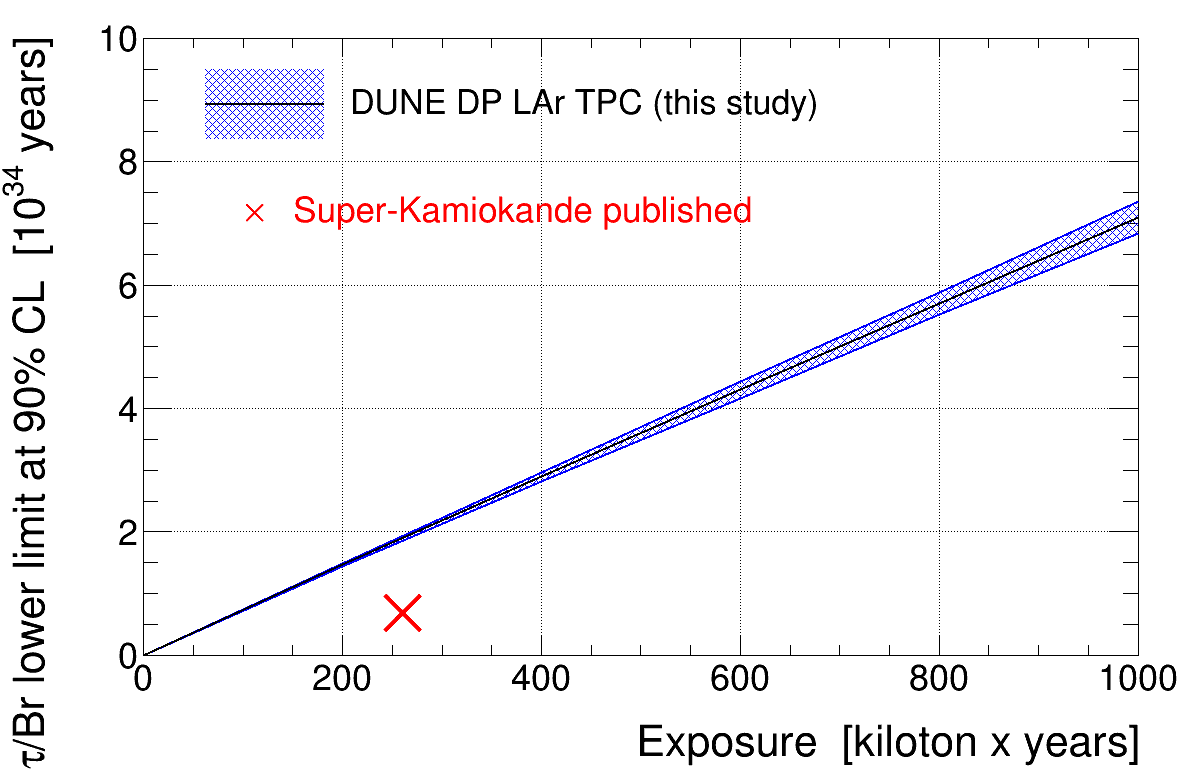}
    \caption{Lower lifetime limit for $\tau/ \text{Br} \left( p \rightarrow \bar{\nu} K^+ \right)$ at $\unit[90]{\%}$ CL as a function of exposure as obtained by the presented study. The black line shows the limit for the mean signal selection efficiencies of both samples and the blue band covers the full spread between the samples. The red cross shows the latest published result by Super-Kamiokande \cite{SuperKPKaon2014}, see section \ref{sec:Intro}.}
    \label{fig:Sensitivity}
\end{figure}

The mean signal selection efficiencies of both samples are used in the sensitivity calculation at the given exposures, and the systematic uncertainty is defined as the full spread between the samples. With $B=0.5$ expected background events and in case no event is observed ($N=0$), the upper limit on the number of signal events at $\unit[90]{\%}$ CL according to Feldman-Cousins is $1.94$ \cite{Feldman:1997qc}. The resulting sensitivities are obtained with equation \ref{eq:sensitivity} and interpolated linearly between the studied exposures, see figure \ref{fig:Sensitivity}. Only the kaon decay mode $K^+ \rightarrow \mu^+ \nu_{\mu}$ has been considered in this study and the obtained results are assumed to be transferable to all other kaon decay modes in the sensitivity calculation, see discussion in section \ref{sec:Discussion}.

The current best published limit of $\tau/ \text{Br} \left( p \rightarrow \bar{\nu} K^+ \right) > \unit[5.9 \cdot 10^{33}]{years}$ by Super-Kamiokande can be reached with an exposure of $\unit[{\sim}80]{kiloton \cdot years}$. After an exposure of 1 megaton $\cdot$ year, a lower limit of \unit[$\tau/ \text{Br} \left( p \rightarrow \bar{\nu}  K^+\right) > 7 \cdot 10^{34}$]{years} can be achieved, reaching the predicted limits of many SUSY GUTs (see section \ref{sec:Intro}).\par
A similar sensitivity study for $p \rightarrow \bar{\nu} K^+$ using a $\unit[{\sim}10]{kiloton}$ single phase LAr TPC at DUNE has been reported in \cite{Abi:2020kei}, reaching a signal selection efficiency of $\unit[15]{\%}$ at a comparable background level. Based on visual scans, reference \cite{Abi:2020kei} further claims that the signal selection efficiency could be increased to $\unit[30]{\%}$ with improved reconstruction algorithms. The signal $K^+$ tracking efficiency in \cite{Abi:2020kei}, which is the share of $K^+$ with a reconstructed track but without any information on the nature of that track, is shown in the left panel of figure \ref{fig:SignalKSelEff} as a function of true kinetic energy. The curve is comparable to the signal $K^+$ selection efficiency after neural network selection in our study, which is the share of $K^+$ that produced a reconstructed track that was already identified as signal $K^+$-like by the neural network. Our analysis benefits from a better charge readout resolution of $\unit[3]{mm}$ compared to $\unit[{\sim}5]{mm}$ in \cite{Abi:2020kei} combined with a dedicated neural network for kaon identification. The charge resolution is important for the detailed reconstruction of the short Bragg peak, which plays a crucial role in the particle identification (see section \ref{sec:KaonID} and figures \ref{fig:EventDisplay}, \ref{fig:StoppingPowerProfiles} and \ref{fig:EventDisplay2}). Another important difference between the dual phase LAr TPC considered in our study and the single phase LAr TPC in \cite{Abi:2020kei} is the number of charge readout views and their orientation: while there are two perpendicular charge readout views that both collect the arriving charge in the dual phase design, the single phase design foresees three charge readout views of which the first two record an induction signal as the charge passes by (induction planes) while the third one collects the arriving charge (collection plane). The angle between the two induction planes and the collection plane is $\unit[\pm 35.7]{^{\circ}}$ \cite{DUNETDRVol4}. The availability of a third readout view can improve the reconstruction and identification of particles that travel parallel to one of the readout views, but no major inefficiencies in particle identification have been found for kaons with such topologies in our study with two perpendicular readout views (see right panel of figure \ref{fig:SignalKSelEff}).

\section{\label{sec:Discussion}Discussion of uncertainties}
The systematic uncertainty related to event generator models was assessed by using two different GENIE tunes. The dominant contribution to the uncertainty originates from the intranuclear propagation of kaons. Although the two tunes use different propagation models with different possible interactions (see section \ref{sec:SignalBkg}), the underlying $K^+$-nucleon scattering cross sections are identical and yield a signal $K^+$ scattering probability of $\unit[32]{\%}$ in both tunes. Since the signal $K^+$ typically lose a large amount of their kinetic energy in the scatters, their tracks are often too short to be identified correctly in the analysis independent of the nature of the scatter. The obtained difference in signal selection efficiencies between the two tunes of about $\unit[2-4]{\%}$ is therefore relatively small, see table \ref{TableSensitivitySignalEff}. Furthermore, the final state interaction rate of $K^+$ inside the remnant nucleus has been cross-checked with NEUT, a generator toolkit developed in the context of the T2K experiment \cite{HayatoNEUT}. NEUT yields a total interaction rate of $\unit[35]{\%}$ with a model combination similar to the alternative GENIE tune, confirming the interaction rate of $\unit[32]{\%}$ obtained with GENIE.\par
The presented analysis was carried out for the most common kaon decay mode $K^+ \rightarrow \mu^+ \nu_{\mu}$ and the obtained results were assumed to be identical for all other kaon decay modes in the sensitivity calculation. The main kaon decay modes and branching ratios are summarized in table \ref{TableKDecayModes}. The event preselection cuts can be easily adjusted for the other kaon decay modes, and the track multiplicity cut 1.3 would likely yield a better background rejection since most background events have a low track multiplicity (see section \ref{sec:EventPreselection} and figure \ref{fig:EventPreselTrackMultiplicity}). Except for $K^+ \rightarrow \pi^+ \pi^+ \pi^-$, which shows more activity at the kaon decay point, the neural network signal $K^+$ track identification is not affected in the remaining decay modes (see section \ref{sec:KaonID}). Subsequently, the presence of multiple particles emerging from the kaon decay point, as well as their correlations, enable a strong background rejection that is expected to be comparable to the one obtained for $K^+ \rightarrow \mu^+ \nu_{\mu}$, and the remaining cuts 4.2 to 4.4 can be adapted accordingly. Moreover, cut 4.2.5, which was introduced for $\mu^+$ traveling in the same direction as the parent $K^+$ and which results in a signal selection efficiency loss of about $\unit[6]{\%}$, is no longer required. It is therefore reasonable to assume that the obtained results for $K^+ \rightarrow \mu^+ \nu_{\mu}$ are transferable to all kaon decay modes in the sensitivity calculation.

\begin{table}[t]
\centering
\begin{tabular}{lc}
\hline
Decay mode & Branching ratio \\ \hline 
\hline
$K^+ \rightarrow \mu^+ \nu_{\mu}$ & $\unit[63.6]{\%}$ \\ 
\hline 
$K^+ \rightarrow \pi^0 e^+ \nu_{e}$ & $\unit[5.1]{\%}$ \\ 
\hline 
$K^+ \rightarrow \pi^0 \mu^+ \nu_{\mu}$ & $\unit[3.4]{\%}$ \\ 
\hline

\end{tabular} \hspace{1cm}
\begin{tabular}{lc}
\hline
Decay mode & Branching ratio \\ \hline 
\hline 
$K^+ \rightarrow \pi^+ \pi^0$ & $\unit[20.7]{\%}$ \\ 
\hline 
$K^+ \rightarrow \pi^+ \pi^+ \pi^-$ & $\unit[5.6]{\%}$ \\ 
\hline 
$K^+ \rightarrow \pi^+ \pi^0 \pi^0$ & $\unit[1.8]{\%}$ \\ 
\hline 
\end{tabular} 
\caption{Main leptonic and semileptonic (left) and hadronic (right) $K^+$ decay modes and branching ratios \cite{PDG}.}
\label{TableKDecayModes}
\end{table}

The detector simulation parameter with the highest impact on the sensitivity limit is the transverse diffusion, see section \ref{sec:DetectorDesignAndSim}. The transverse smearing of the charge at the starting and stopping points of a particle reduces the reconstructed charge in the first and last hit and makes the particle's reconstructed track appear longer and tilted. These effects lead to a smearing of the $\langle -dE/ds \rangle$ vs. $E_\text{kin, residual}$ stopping power profiles used for the neural-network-driven signal $K^+$ track identification, which plays a central role in the presented analysis. Since the mean transverse displacement $\lambda_\text{T}$ is proportional to $\sqrt{t_\text{Drift}}$, placing all events in the center of the detector at \unit[6]{m} drift in this analysis effectively leads to a higher mean transverse displacement and therefore to a bigger smearing of the $\langle -dE/ds \rangle$ vs. $E_\text{kin, residual}$ curves compared to the expected random distribution of events between 0 and \unit[12]{m} drift.\par
A process for charged kaon production in atmospheric neutrino interactions on nuclei that has not been considered in this study is the so-called charged current coherent $K^+$ production, in which the neutrino scatters off the entire argon nucleus to create an on-shell $K^+$ while leaving the nucleus intact. First evidence for this process has recently been found by the MINERvA experiment \cite{MINERVACoherentKaon}. Although no particles leave the nucleus, the lepton from the charged current interaction makes this process distinguishable from the signal. Cosmic muon-induced backgrounds for $p \rightarrow \bar{\nu} K^+$ have been found to be negligible for large rock overburdens in our previous study and were therefore not considered in this analysis \cite{ABueno07}.

\section{\label{sec:Conclusions}Conclusions}
We have used the $p \rightarrow \bar{\nu} K^+$ benchmark channel to update our previously found sensitivity limits for several proton and neutron decay modes. In our previous study, we performed a simplified detector simulation and had to make assumptions on the detector and backgrounds. Since then, a well-defined DP LAr TPC detector design has been established, precision neutrino cross section measurements have been carried out and more sophisticated event generators have become available. These developments allowed us to update our results for the proton decay mode $p \rightarrow \bar{\nu} K^+$ with a full detector simulation and improved signal and atmospheric neutrino background samples. In this study, we have found a signal selection efficiency of $\unit[{\sim}50]{\%}$ in quasi-background-free conditions, resulting in a lower lifetime limit of \unit[$\tau/ \text{Br} \left( p \rightarrow \bar{\nu}  K^+\right) > 7 \cdot 10^{34}$]{years} at \unit[90]{$\%$} CL for an exposure of \unit[1]{megaton $\cdot$ year}.\par
The decrease in signal selection efficiency with respect to the $\unit[{\sim}97]{\%}$ found in our previous study can largely be explained by a low signal $K^+$ identification efficiency for low-energy $K^+$ that scattered inside the nucleus and by badly reconstructed tracks with difficult topologies, especially parallel or anti-parallel to the drift direction, two effects that have previously not been considered to their full extent (see figure \ref{fig:SignalKSelEff}). A better spatial resolution could improve the reconstruction of low-energy $K^+$ and increase the sensitivity to proton decay via $p \rightarrow \bar{\nu} K^+$.\par
While the detector design and simulation parameters are well defined, the reconstruction and analysis used in this study can be further improved to yield a higher signal selection efficiency, which is supported by the fact that the event displays of some of the most persistent background events in this analysis are clearly distinguishable from the signal (see figures \ref{fig:EventDisplay} and \ref{fig:EventDisplay2}). An aided pattern recognition was used in this study instead of a full pattern recognition algorithm (see section \ref{sec:Reco}), but the additional loss in signal selection efficiency by using a full pattern recognition algorithm is expected to be small since it would mainly affect events with short tracks and difficult topologies that already failed the selection cuts in the presented analysis. Except for said short tracks, the neural network signal $K^+$ identification shows a good performance with losses of only $\unit[{\sim}5]{\%}$ for kaons above $\unit[80]{MeV}$ (see left panel of figure \ref{fig:SignalKSelEff}).\par
Considering the latest published Super-Kamiokande result with a signal selection efficiency of $\epsilon \lesssim \unit[10]{\%}$ and ${\sim}0.5$ expected background events at an exposure of \unit[260]{kiloton $\cdot$ years} for $p \rightarrow \bar{\nu} K^+$ \cite{SuperKPKaon2014}, we can confirm that the LAr TPC technology is superior over Water Cherenkov detectors for many of the challenging nucleon decay modes. Moreover, LAr TPCs are ideal for discoveries at the few events level thanks to their excellent imaging capabilities and concomitant background rejection.

\acknowledgments
This work was supported by the Swiss National Science Foundation (grant number SNSF 200020\_188533) and ETH Z\"urich. We would like to acknowledge the development of the oscillation code for the HKKM2014 atmospheric neutrino flux by Ivan Martinez Soler and the technical help with GENIE by Stephen Dolan and Joshua L. Barrow.

\end{document}